\documentclass[twoside]{IEEEtran}
\usepackage[caption=false,font=footnotesize]{subfig}
\usepackage{xcolor}    
\usepackage{amsmath}   
\allowdisplaybreaks[4]
\usepackage{amssymb}   
\usepackage{amsfonts}  
\usepackage{mathrsfs}  
\usepackage{array,bm}  
\usepackage{textcomp}  
\usepackage{stfloats}  
\usepackage{verbatim}  
\usepackage{graphicx}  
\usepackage{cite,url}  
\usepackage{booktabs}  
\usepackage{multirow}  
\usepackage{lettrine}  
\usepackage{mathtools} 
\usepackage{orcidlink} 
\usepackage[ruled,linesnumbered]{algorithm2e} %
\usepackage{lipsum}
\newtheorem{lemma}{\bf{Lemma}}
\newtheorem{proposition}{\bf{Proposition}}
\graphicspath{{pic/}{Fig/}{paperTex/pic/}{paperTex/Fig/}} 
\usepackage{hyperref}  
\begin{document}
\bstctlcite{BSTcontrolFirstAuthor}

\title{\huge Movable Antenna-Assisted Covert Communications: Robust Beamforming and Antenna Position Optimization}

\author{
  Xiangyu Dong$^{\orcidlink{0009-0004-4606-2271}}$,
  Ran Yang$^{\orcidlink{0000-0002-0435-7926}}$,
  Weidong Mei$^{\orcidlink{0000-0002-8113-4283}}$,~\IEEEmembership{Member,~IEEE},
  Zheng Dong$^{\orcidlink{0000-0002-9224-324X}}$,~\IEEEmembership{Member,~IEEE},
  Yue Xiu$^{\orcidlink{0000-0002-8433-6497}}$,\\
  Daniel Benevides da Costa$^{\orcidlink{0000-0002-5439-7475}}$,~\IEEEmembership{Senior Member,~IEEE},
  Zhongpei Zhang$^{\orcidlink{0000-0003-2772-9937}}$,
  Ning Wei$^{\orcidlink
  {0000-0002-3936-7526}}$
  \vspace{-1em}
  \thanks{
  This work was supported in part by the National Key Research and Development Program of China under Grant 2024YFB2907900, and in part by Shenzhen Natural Science Foundation under Grant JCYJ20250604180459039.

  Xiangyu Dong, Weidong Mei, Ran Yang, Yue Xiu and Ning Wei are with the National Key Laboratory of Wireless Communications, University of Electronic Science and Technology of China, Chengdu 611731, China (e-mail: wmei@uestc.edu.cn).

  Zheng Dong is with the School of Information Science and Engineering, Shandong University, Qingdao 266237, China (e-mail: {\urlstyle{same}\nolinkurl{zhengdong@sdu.edu.cn}}).

  Daniel Benevides da Costa is with the Interdisciplinary Research Center for Communication Systems and Sensing (IRC-CSS), Department of Electrical Engineering, King Fahd University of Petroleum and Minerals (KFUPM), Dhahran 31261, Saudi Arabia (e-mail danielbcosta@ieee.org).

  Zhongpei Zhang is with Shenzhen Institute for Advanced Study, University of Electronic Science and Technology of China, Shenzhen 518110, China (e-mail: zhangzp@uestc.edu.cn).
}
}
\maketitle

\begin{abstract}
In this paper, we study a movable antenna (MA)-enhanced covert communication system, where a legitimate transmitter (named Alice) serves multiple legitimate users in the presence of multiple aerial wardens that attempt to detect the transmission from Alice.
To ensure data covertness with imperfect knowledge of warden channel state information (CSI), we develop a unified robust covert transmission framework that incorporates bounded and statistical CSI error models.
In particular, we aim to jointly optimize the transmit beamforming, the artificial noise (AN) covariance matrix, and MA positions at Alice to maximize the minimum weighted signal-to-interference-plus-noise ratio (SINR) among the legitimate users subject to data covertness constraints.
The resulting optimization problem is intractable to solve optimally due to the strong coupling among the optimization variables and the random channel uncertainty.
To deal with this issue, we propose a block coordinate descent (BCD)-based algorithm.
Specifically, for the norm-bounded CSI error model, we first employ the S-procedure to convert the covertness constraints into a sequence of linear matrix inequalities (LMIs). Then, a successive convex approximation (SCA) procedure incorporating semidefinite relaxation (SDR) and bisection search techniques is proposed to solve the resulting problem.
For the statistical CSI error model, the Bernstein-type inequality is employed to derive deterministic approximations of the probabilistic constraint in terms of LMIs and second-order cone (SOC) constraints that can be efficiently tackled.
Numerical results are provided to demonstrate the effectiveness and robustness of the proposed algorithm in improving the performance of MA-enhanced covert communication systems.
\end{abstract}


\setlength{\abovedisplayskip}{2pt plus 1pt minus 1pt}
\setlength{\belowdisplayskip}{2pt plus 1pt minus 1pt}
\setlength{\abovedisplayshortskip}{1pt plus 1pt}
\setlength{\belowdisplayshortskip}{1pt plus 1pt}
\setlength{\jot}{1pt}
\setlength{\textfloatsep}{0.1cm}
\setlength{\skip\footins}{2pt plus 1pt minus 1pt}
\makeatletter
\def\subsection{\@startsection{subsection}{2}{\z@}{0.6ex plus 0.4ex minus 0.2ex}%
{0.7ex plus .5ex minus 0ex}{\normalfont\normalsize\itshape}}%
\makeatother

\section{Introduction}\label{sec:intro}
\IEEEPARstart{T}{he} sixth-generation (6G) mobile networks are envisioned to provide ubiquitous connectivity for emerging applications, including autonomous vehicles, smart factories, and embodied intelligence~\cite{Vision6G}.
Conventional wireless networks primarily serve terrestrial devices, whereas 6G networks are expected to integrate terrestrial and aerial networks.
In particular, the low-altitude economy (LAE) is driving the deployment of a growing number of uncrewed aerial vehicles (UAVs) \cite{LAENet}. 
These UAVs can be integrated into cellular networks as aerial users~\cite{aerialuser} or aerial sensing platforms~\cite{MAISAC2} for applications such as surveillance, data collection, and aerial filming \cite{ZhanUAVSensing}. 
However, the broadcast nature of wireless channels and the high-quality links associated with aerial devices expose low-altitude transmissions to security risks such as eavesdropping, information leakage, and jamming \cite{AlHabobAerialPLS,XiuMAAntiJammingMEC}. 
These risks underscore the fundamental need for reliable and secure wireless transmission in low-altitude networks.

Physical-layer security (PLS) approaches leverage differences between legitimate and eavesdropping channels to reduce information leakage and enhance the secrecy capacity of wireless systems \cite{Wyner1975}.
Existing PLS techniques mainly include artificial noise (AN) \cite{DongFCLASecure}, directional modulation (DM) \cite{YangDiscreteRISSLP}, and null-space beamforming \cite{Qian6DMASecure2026}.
These PLS techniques have been widely adopted in various wireless communication scenarios, such as simultaneous wireless information and power transfer (SWIPT) \cite{DongMASecureSWIPT}, integrated sensing and communication (ISAC) \cite{XiuRISAAVISAC}, physical-layer service integration \cite{citeformeipls}, and airborne communications \cite{ZhuToMA,XiuSecureISACAAV2025}.
Although these measures can protect the confidentiality of transmitted message, they do not necessarily conceal the existence of a transmission, thereby increasing the risk of exposing transmission behavior.

In this context, covert communication, also known as low probability of detection (LPD) communication, has drawn wide
attention for its capability to conceal transmission behavior \cite{WangPLSCovert}.
The authors in \cite{Bash2013} pioneered to establish the fundamental square root law, which states that $\mathcal{O}(\sqrt{N})$ bits can be delivered reliably and covertly over $N$ uses of an additive white Gaussian noise (AWGN) channel.
In \cite{SobersJammer2017}, an uninformed jammer transmitting AN was proposed to increas the number of reliably and covertly transmitted bits from $\mathcal{O}(\sqrt{N})$ to $\mathcal{O}(N)$.
The authors in \cite{SoltaniAN} investigated a random network in which the friendly transmitter nearest the warden generates AN.
Covert communication has also been investigated in air-ground networks.
The authors in \cite{WangUAVSurveillance} investigated multi-hop secrecy and covert communication between terrestrial nodes under UAV surveillance.
In \cite{ChenUAVFlyingWarden}, a UAV relay was investigated to supported finite-blocklength covert communication against a flying warden.

In particular, multi-antenna techniques have been widely adopted in covert communications by offering spatial degrees of freedom (DoFs) in both detection and transmission.
From the detection perspective, the authors in \cite{HuBeamSweeping} derived the optimal number of beam-sweeping sectors that minimizes the detection error probability (DEP) of a UAV transmission.
For covert transmission, centralized and distributed multi-antenna architectures were proposed to improve covert throughput in random wireless networks \cite{ZhengMultiAntennaCovert}.
The authors in \cite{ForouzeshNull} combined three-dimensional transmit beamforming with multi-antenna jamming nulled at the intended receiver to improve the covert rate.
Although these optimization schemes have demonstrated the benefits of multi-antenna transmission, most prototypes and algorithm development are based on conventional fixed-position antenna (FPA) arrays. 
Such fixed array geometry leaves the spatial DoFs associated with antenna positions unexploited, leading to limited covertness performance improvement.
More importantly, when intended receiver and warden channels are highly correlated, the spatial discrimination capability becomes insufficient, thereby cannot strengthen the desired links while  effectively suppressing information leakage.

Recently, movable antennas (MAs) have emerged as an innovative solution to overcome this limitation by dynamically adjusting the antenna positions within a wavelength-scale region~\cite{MA4,YangFlexibleWMMSE2025}.
Unlike conventional FPA arrays, MA arrays can avoid deep fading and reshape the composite channel through changes in the propagation phases of the multipath components with antenna repositioning in real time~\cite{MA1}.
In particular, antenna movement can be realized using mechanically driven, liquid-based, or electronically reconfigurable architectures.
For instance, the authors in \cite{WeiMechanicalPower2026} developed a power consumption model for motor-driven MA systems by jointly optimizing the antenna positions, moving speeds, and transmit precoding to maximize energy efficiency.
The authors in \cite{DongMAPrototype} developed a motor-driven MA prototype with a slide track positioning accuracy of 0.05~mm and experimentally verified the received power gains of antenna movement at 3.5 and 27.5~GHz.
The ability of MA arrays to improve channel conditions has motivated further investigation of the benefits of jointly optimizing the MA positions and beamforming~\textcolor{red}{\cite{YangRobustRISMA2026}}.
For example, the authors in \cite{WangMNFBeamCoverage} used Fourier transform theory to construct a spatial multi-notch filter and select MA positions from a discrete set of sampling points for multi-region beam coverage.
The same filter-based view was applied to multi-beam forming by reducing the pass bands over continuous angular regions to impulses at the desired discrete angles~\cite{MeiGeneralAPO2026}.
Under the derived feasibility condition, the authors in \cite{zlp1} analytically showed that joint MA position and beamforming optimization can achieve full array gain in the desired direction while forming nulls in all specified undesired directions.
Furthermore, the authors in \cite{citeformeirobust} unveiled that antenna position optimization can reduce the performance loss due to imperfect CSI.

The aforementioned studies provided a theoretical basis for using MA arrays to suppress signal leakage toward undesired directions.
This ability has also been exploited in covert communication.
For example, the authors in \cite{WangWavelengthCovert} attained wavelength-scale beamforming resolution by adjusting the antenna positions, which preserves transmission covertness even when the warden is close to the intended receiver.
In \cite{LiuMACovert}, the transmit beamforming and MA positions were jointly optimized for a single covert user in the presence of multiple non-colluding wardens.
The authors in \cite{MaoMultiuserCovert} maximized the sum rate of multiple covert users in the presence of a single warden.
In \cite{ChengMFDACovert}, a movable frequency diverse array was adopted against multiple wardens with uncertain locations.
The authors in \cite{HuFDMACovert} employed a full-duplex receiver with MAs to transmit AN against an MA-equipped warden.
Although these works have demonstrated the effectiveness of MAs, they are typically based on the perfect knowledge of the warden channel state information (CSI).
In practice, perfect CSI of non-cooperative wardens is challenging for Alice to acquire~\cite{ForouzeshCDI,CovertSurvey}.
Therefore, ignoring the inherent CSI error may violate the covertness requirement, leading to degraded security performance~\cite{ZhouMixedFieldCovert}.
To address this issue, the authors in \cite{YangMAISACC2026} investigated robust beamforming and antenna position optimization for covert ISAC systems.
However, the proposed approach in \cite{YangMAISACC2026} was limited to scenarios with a single warden and cannot be applied to other error models beyond the bounded error model, i.e., the channel error is bounded by a prescribed value.
As presented in Table~\ref{tab:related_work_comparison}, the proposed framework is compared with related studies in terms of system models, CSI assumptions, joint optimization, and covertness guarantees.
To enhance data security and further unlock the synergy gain between MAs and covert transmission, a unified robust MA-based transmission framework is highly imperative. 

\begin{table*}[t]
\centering
\caption{Comparison with closely related covert communication studies.}
\label{tab:related_work_comparison}
\small
\setlength{\tabcolsep}{3pt}
\renewcommand{\arraystretch}{1.22}
\begin{tabular}{|>{\centering\arraybackslash}p{0.38\textwidth}|*{8}{>{\centering\arraybackslash}p{0.05\textwidth}|}>{\centering\arraybackslash}p{0.09\textwidth}|}
\hline
 & \cite{WangWavelengthCovert} & \cite{LiuMACovert} & \cite{MaoMultiuserCovert} & \cite{ChengMFDACovert} & \cite{HuFDMACovert} & \cite{ZhouMixedFieldCovert} & \cite{YangMAISACC2026} & \cite{ZhangRobustCovertISAC} & \textbf{Proposed} \\
\hline
Multiple wardens & & $\checkmark$ & & $\checkmark$ & & $\checkmark$ & & $\checkmark$ & $\checkmark$ \\
\hline
Multiple overt users & & & & & & & & $\checkmark$ & $\checkmark$ \\
\hline
Multiple covert users & & & $\checkmark$ & & & $\checkmark$ & $\checkmark$ & $\checkmark$ & $\checkmark$ \\
\hline
Imperfect warden channel information & & & & $\checkmark$ & $\checkmark$ & & $\checkmark$ & $\checkmark$ & $\checkmark$ \\
\hline
Joint AN optimization & & & & & $\checkmark$ & & & & $\checkmark$ \\
\hline
MA position optimization & $\checkmark$ & $\checkmark$ & $\checkmark$ & $\checkmark$ & $\checkmark$ & $\checkmark$ & $\checkmark$ & & $\checkmark$ \\
\hline
Worst-case CSI-error guarantee & & & & & & & $\checkmark$ & $\checkmark$ & $\checkmark$ \\
\hline
Covertness outage guarantee & & & & & & & & $\checkmark$ & $\checkmark$ \\
\hline
\end{tabular}
\end{table*}

In this paper, we study robust MA-assisted covert communications with multiple wardens under two typical warden CSI error models. The main contributions are summarized as follows.
\begin{itemize}[\setlength{\itemsep}{0pt plus 1pt}]
\item We introduce an MA-assisted covert communication system with Alice serving legitimate users in the presence of multiple wardens. To guarantee transmission covertness, we develop a unified optimization framework to maximize the weighted max-min signal-to-interference-plus-noise ratio (SINR) by jointly optimizing transmit beamforming, AN covariance matrix, and antenna placement, subject to a transmit power budget and a covertness constraint at each warden. 
The nonlinear coupling between MA positions and uncertain warden channels complicates the derivation of tractable constraint approximations that preserve the required covertness guarantees.
In addition to norm-bounded CSI errors, statistical warden CSI errors are also accounted for, assuming that they follow complex Gaussian distribution.
\item For the norm-bounded errors, we develop a worst-case optimization scheme that enforces the covertness constraints for every error realization in the channel uncertainty set. 
To address the challenging problem, we first apply the S-procedure to reformulate the intractable semi-infinite quadratic condition into an equivalent linear matrix inequality (LMI). 
Then, we develop a block coordinate descent (BCD) algorithm that integrates semidefinite relaxation (SDR) and bisection search for the covariance matrix optimization with a quadratic transform and successive convex approximation (SCA) for the position optimization.
\item For the statistical CSI error, we propose an outage-constrained optimization scheme to ensure covertness with a prescribed probability guarantee. 
A deterministic approximation of the covertness constraint is derived by using the Bernstein-type inequality (BTI) for complex Gaussian quadratic forms and is expressed as LMI and second-order cone (SOC) constraints that can be efficiently tackled by using the proposed BCD optimization framework.
\item Simulation results demonstrate the effectiveness of the proposed schemes and provide insights into robust MA-assisted covert transmission.
In particular, the proposed MA schemes outperform the conventional FPA schemes in terms of weighted max-min SINR under both CSI error models, with further improvements observed for larger movable regions and richer multipath propagation.
Moreover, the statistical optimization scheme outperforms the worst-case scheme by allowing a prescribed covertness outage probability.
\end{itemize}

\section{System Model}\label{sys_model}
\subsection{Signal Model}
\begin{figure}[t]
  \centering
  \includegraphics[width=\columnwidth]{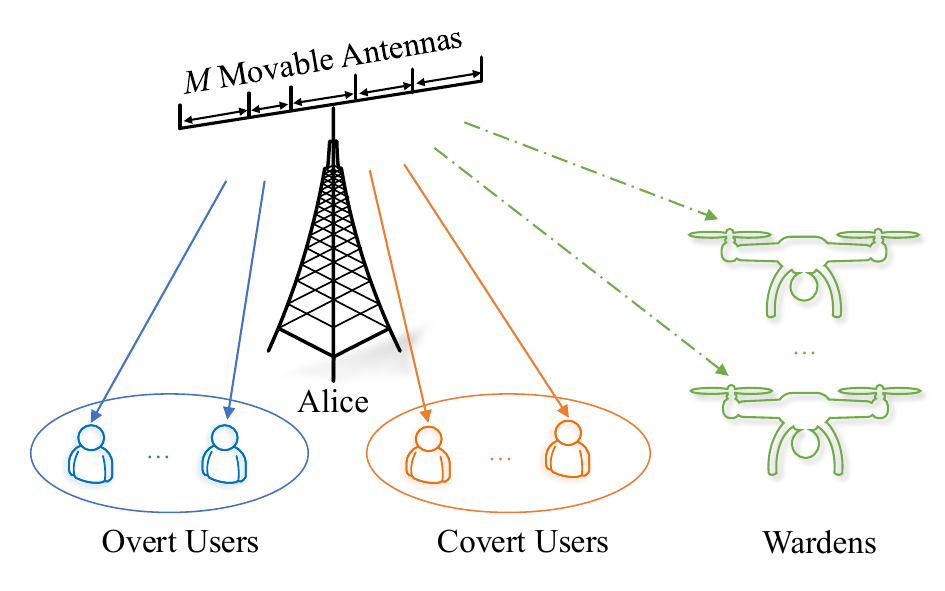}
  \caption{MA-assisted covert communication system.}
  \label{fig:system_model}
\end{figure}

We consider a downlink multiuser multiple-input single-output (MU-MISO) covert communication system, as shown in Fig.~\ref{fig:system_model}.
It is assumed that Alice is equipped with $M$ MAs, and serves $K=K_o+K_c$ single-antenna legitimate users in the presence of $Q$ single-antenna wardens.
Let $\mathcal{M}\triangleq\{1,\ldots,M\}$ denote the set of transmit MA indices.
The legitimate users are divided into $K_o$ overt users and $K_c$ covert users. 
The overt signals can be exploited as a cover to enable the covert transmission of confidential information.
The set of overt users is denoted by $\mathcal{K}_o = \{1, \ldots, K_o\}$, the set of covert users is denoted by
$\mathcal{K}_c = \{K_o+1, \ldots, K_o+K_c\}$, and the set of all legitimate users is $\mathcal{K} = \mathcal{K}_o \cup \mathcal{K}_c$.
The set of wardens is $\mathcal{Q}=\{1,\dots,Q\}$. Each warden is a detector carried by a rotary-wing UAV that hovers over the served area, and it passively observes its received signal and tests whether a covert transmission is present. 

Let $\mathcal{H}_1$ and $\mathcal{H}_0$ denote the hypotheses that Alice does and does not transmit the covert signal, respectively. Under the two hypotheses, the signal transmitted by Alice is given by
\begin{equation}\label{eq:tx_signal}
\left\{
\begin{aligned}
\mathcal{H}_1 :\;\boldsymbol{s}[n] &= \boldsymbol{W}_o \boldsymbol{c}_o[n] + \boldsymbol{W}_c \boldsymbol{c}_c[n] + \boldsymbol{r}[n],\\
\mathcal{H}_0 :\;\boldsymbol{s}[n] &= \boldsymbol{W}_o \boldsymbol{c}_o[n] + \boldsymbol{r}[n],
\end{aligned}
\right.
\end{equation}
where $\boldsymbol{s}[n] \in \mathbb{C}^{M\times 1}$ is the transmit signal in the $n$-th time slot, $n\in\{1,\ldots,N\}$, with $N$ being the transmission blocklength, $\boldsymbol{W}_o = [\boldsymbol{w}_1, \ldots, \boldsymbol{w}_{K_o}] \in \mathbb{C}^{M\times K_o}$ and $\boldsymbol{W}_c = [\boldsymbol{w}_{K_o+1}, \ldots, \boldsymbol{w}_{K_o+K_c}] \in \mathbb{C}^{M\times K_c}$ are the beamforming matrices for the overt and covert users, respectively, $\boldsymbol{c}_o[n] \in \mathbb{C}^{K_o\times 1}$ and $\boldsymbol{c}_c[n] \in \mathbb{C}^{K_c\times 1}$ are the corresponding overt and covert data streams whose entries are independent and identically distributed as $\mathcal{CN}(0,1)$, and $\boldsymbol{r}[n] \sim \mathcal{CN}(\bm{0}_M, \boldsymbol{R}_{\rm A})$ is the AN vector with covariance matrix $\boldsymbol{R}_{\rm A} \succeq \boldsymbol{0}$.
For brevity, we define $\boldsymbol{W}\triangleq[\boldsymbol{W}_o,\boldsymbol{W}_c]$. 

\subsection{Channel Model}
We assume that the MAs can move within a linear region $\mathcal{X}=[0,A]$, where $A$ denotes the region length. The position of the $m$-th MA is denoted by $x_m\in\mathcal{X}$, and the antenna position vector (APV) is defined as $\boldsymbol{x}=[x_1,\ldots,x_M]^T\in\mathbb{R}^{M\times1}$.
We consider a far-field channel model in which the size of the moving region is significantly smaller than the signal propagation distance~\cite{P_FAA}. 
Thus, the antenna movement does not affect the angles of departure (AoDs) \cite{zlp2}.
Let $L_k$ denote the number of channel paths between Alice and legitimate user $k$. The azimuth AoD of the $\ell$-th path to user $k$ is denoted by $\theta_{k}^{\ell} \in [0,\pi]$, where $1 \leq \ell \leq L_k$.
The signal propagation distance difference between the $m$-th transmitting MA position $x_m$ and the reference point $\boldsymbol{o}^t$ is given by
$\chi_{k}^{\ell}(x_m) = x_m\cos\theta_{k}^{\ell}$.
To incorporate the phase
differences across all $L_k$ transmit paths, the field response vector (FRV) at $x_m$ is then given by
\begin{equation}
\boldsymbol{g}_{k}(x_m) =
  \left[
    e^{\jmath\frac{2\pi}{\lambda}\chi^{1}_{k}(x_m)},\cdots,
    e^{\jmath\frac{2\pi}{\lambda}\chi^{L_k}_{k}(x_m)}
  \right]
  ^T \in \mathbb{C}^{L_k\times 1},
\end{equation}
where $\lambda$ is the wavelength of the carrier signal.
As a result, the field response matrix (FRM) from Alice to user $k$ over all transmit MAs is
\begin{equation}
\boldsymbol{G}_k(\boldsymbol{x}) =
  \left[
    \boldsymbol{g}_{k}(x_1), \cdots, \boldsymbol{g}_{k}(x_M)
  \right] \in \mathbb{C}^{L_k\times M}.
\end{equation}
Let $\boldsymbol{f}_k = [\beta_{k}^{*,1},\cdots,\beta_{k}^{*,L_k}]^T \in \mathbb{C}^{L_k\times 1}$ be the path response vector (PRV) from Alice to the $k$-th user. Then, the channel vector at the $k$-th user is given by
\begin{equation}\label{eq:user_channel}
  \boldsymbol{h}_k^H(\boldsymbol{x}) = \boldsymbol{f}_k^H\boldsymbol{G}_k(\boldsymbol{x}) \in \mathbb{C}^{1\times M},\forall k\in\mathcal{K}.
\end{equation}

As the warden is assumed to be a UAV, a line-of-sight (LoS) model is adopted for the warden channel, i.e.,
\begin{equation}\label{eq:warden_channel}
\boldsymbol{h}_{q}^H(\boldsymbol{x}) = \beta_q \, \boldsymbol{g}_{q}^H(\boldsymbol{x}), \forall q \in \mathcal{Q},
\end{equation}
where $\beta_q$ is the LoS complex gain between warden $q$ and Alice with AoD $\theta_{q}$, and $\boldsymbol{g}_{q}^H(\boldsymbol{x})$ is the corresponding FRV given by
\begin{equation}\label{eq:warden_channel_FRV}
\boldsymbol{g}_{q}^H(\boldsymbol{x}) =
  \left[
    e^{\jmath\frac{2\pi}{\lambda}\chi_{q}(x_1)},\cdots,
    e^{\jmath\frac{2\pi}{\lambda}\chi_{q}(x_M)}
  \right] \in \mathbb{C}^{1\times M},
\end{equation}
with $\chi_{q}(x_m) = x_m\cos\theta_{q}$ denoting the propagation distance difference between the $m$-th transmitting MA position and the reference point $\boldsymbol{o}^t$ for warden $q$.

\subsection{Communication Performance}

The received signal at the $k$-th legitimate user under $\mathcal{H}_1$ is given by
\begin{equation}\label{eq:rx_user}
\begin{aligned}
y_k[n] = 
&\underbrace{\boldsymbol{h}_k^H(\boldsymbol{x})\, \boldsymbol{w}_k c_k[n]}_{\text{desired signal}} 
+ \underbrace{\sum_{i\in\mathcal{K}\setminus\{k\}} \boldsymbol{h}_k^H(\boldsymbol{x})\, \boldsymbol{w}_i c_i[n]}_{\text{multiuser interference}}
\\&
+ \underbrace{\boldsymbol{h}_k^H(\boldsymbol{x})\, \boldsymbol{r}[n]}_{\text{AN interference}}
+ z_k[n],
\end{aligned}
\end{equation}
where $z_k[n] \sim \mathcal{CN}(0,\sigma_k^2)$ denotes the AWGN at the $k$-th legitimate user.
The corresponding SINR at the $k$-th legitimate user under $\mathcal{H}_1$ is given by
\begin{equation}\label{eq:sinr_user}
\begin{aligned}
&\gamma_{1,k}^C(\boldsymbol{W},\boldsymbol{R}_{\rm A},\boldsymbol{x}) = \\&
\frac{\boldsymbol{h}_k^H(\boldsymbol{x}) \boldsymbol{w}_k \boldsymbol{w}_k^H \boldsymbol{h}_k(\boldsymbol{x})}
{\sum_{i\in\mathcal{K}\setminus\{k\}} \!\boldsymbol{h}_k^H(\boldsymbol{x}) \boldsymbol{w}_i \boldsymbol{w}_i^H \boldsymbol{h}_k(\boldsymbol{x})
\!+\! \boldsymbol{h}_k^H(\boldsymbol{x}) \boldsymbol{R}_{\rm A}\boldsymbol{h}_k(\boldsymbol{x}) \!+\! \sigma_k^2}.
\end{aligned}
\end{equation}

Likewise, the SINR of overt user $k\in\mathcal K_o$ under $\mathcal{H}_0$ is given by\footnote{Note that due to the additional interference from the covert users that is present under $\mathcal{H}_1$ but absent under $\mathcal{H}_0$, we can verify that $\gamma_{1,k}^C(\boldsymbol{x}) \le \gamma_{0,k}^C(\boldsymbol{x})$ for all $k\in\mathcal K_o$.}
\begin{equation}
\begin{aligned}
&\gamma_{0,k}^C(\boldsymbol{W},\boldsymbol{R}_{\rm A},\boldsymbol{x}) = \\&
\frac{\boldsymbol{h}_k^H(\boldsymbol{x}) \boldsymbol{w}_k \boldsymbol{w}_k^H \boldsymbol{h}_k(\boldsymbol{x})}
{\sum_{i\in\mathcal{K}_o\!\setminus\!\{k\}}\! \boldsymbol{h}_k^H(\boldsymbol{x}) \boldsymbol{w}_i \boldsymbol{w}_i^H \boldsymbol{h}_k(\boldsymbol{x})
\!+\! \boldsymbol{h}_k^H(\boldsymbol{x}) \boldsymbol{R}_{\rm A}\boldsymbol{h}_k(\boldsymbol{x}) \!+\! \sigma_k^2}.
\end{aligned}
\end{equation}
\subsection{Detection Performance and Covertness Constraint}
We consider a non-colluding warden scenario, where each warden independently detect the communication behaviour between Alice and the covert users based on the Neyman--Pearson test~\cite{ZhouMixedFieldCovert}. 
Specifically, the received signal at warden $q$ in time slot $n$ can be given by
\begin{equation}\label{eq:rx_warden}
\left\{
\begin{aligned}
\mathcal{H}_1: y_{q}[n] & = \boldsymbol{h}_{q}^H(\boldsymbol{x}) \big( \boldsymbol{W}_o \boldsymbol{c}_o[n] + \boldsymbol{W}_c \boldsymbol{c}_c[n] + \boldsymbol{r}[n] \big) + z_{q}[n], \\
\mathcal{H}_0: y_{q}[n] & = \boldsymbol{h}_{q}^H(\boldsymbol{x}) \big( \boldsymbol{W}_o \boldsymbol{c}_o[n] + \boldsymbol{r}[n] \big) + z_{q}[n],
\end{aligned}%
\right.
\end{equation}%
where $z_{q}[n] \sim \mathcal{CN}(0, \sigma_{q}^2)$ denotes the AWGN at warden $q$.
The signals received at warden $q$ over the $N$ time slots are stacked into the observation vector $\boldsymbol{y}_{q} = [y_{q}[1], \ldots, y_{q}[N]]^T \in \mathbb{C}^{N\times 1}$.
Given that the received samples are independent and identically distributed across time slots, the probability density functions (PDFs) of $\boldsymbol{y}_{q}$ under two hypotheses are respectively given by
\begin{equation}\label{eq:warden_pdf1}
\mathbb{P}\left(\boldsymbol{y}_{q}|\mathcal{H}_1\right) \!=\! \frac{\exp\!\left(- \frac{\|\boldsymbol{y}_{q}\|^2}{\Lambda_{1,q}(\boldsymbol{x})}\right)}{\pi^N \Lambda_{1,q}^N(\boldsymbol{x})},
\mathbb{P}\left(\boldsymbol{y}_{q}|\mathcal{H}_0\right) \!=\! \frac{\exp\!\left(- \frac{\|\boldsymbol{y}_{q}\|^2}{\Lambda_{0,q}(\boldsymbol{x})}\right)}{\pi^N \Lambda_{0,q}^N(\boldsymbol{x})},
\end{equation}
where
$
\Lambda_{1,q}(\boldsymbol{x}) \triangleq \boldsymbol{h}_{q}^H(\boldsymbol{x}) \!\left(\sum_{k\in\mathcal{K}} \boldsymbol{w}_k\boldsymbol{w}_k^H + \boldsymbol{R}_{\rm A}\right) \boldsymbol{h}_{q}(\boldsymbol{x}) + \sigma_{q}^2$, 
and $\Lambda_{0,q}(\boldsymbol{x}) \triangleq \boldsymbol{h}_{q}^H(\boldsymbol{x}) \!\left(\sum_{k\in\mathcal{K}_o} \boldsymbol{w}_k\boldsymbol{w}_k^H + \boldsymbol{R}_{\rm A}\right) \boldsymbol{h}_{q}(\boldsymbol{x}) + \sigma_{q}^2$.

Denote by $\mathcal{D}_{1,q}$ and $\mathcal{D}_{0,q}$ the binary decisions of warden $q$ in support of $\mathcal{H}_1$ and $\mathcal{H}_0$, respectively.
Following the work in \cite{Ran1}, we assume that each warden uses classical hypothesis testing with equal prior probabilities for the two hypotheses.
The DEP at warden $q$ is then given by $\xi_q = \mathbb{P}_{{\rm MD},q} + \mathbb{P}_{{\rm FA},q}$, where
$\mathbb{P}_{{\rm MD},q} \triangleq \mathbb{P}\left(\mathcal{D}_{0,q}|\mathcal{H}_1\right)$ and
$\mathbb{P}_{{\rm FA},q} \triangleq \mathbb{P}\left(\mathcal{D}_{1,q}|\mathcal{H}_0\right)$
are the missed detection and false alarm probabilities, respectively.
To minimize the DEP $\xi_q$, warden $q$ performs the likelihood ratio test between $\mathbb{P}\left(\boldsymbol{y}_{q}|\mathcal{H}_1\right)$ and $\mathbb{P}\left(\boldsymbol{y}_{q}|\mathcal{H}_0\right)$.
The likelihood ratio test therefore compares the received power with a threshold, i.e.,
\begin{equation}\label{eq:LRT}
\frac{\mathbb{P}\left(\boldsymbol{y}_{q}|\mathcal{H}_1\right)}{\mathbb{P}\left(\boldsymbol{y}_{q}|\mathcal{H}_0\right)}
\underset{\mathcal{D}_{0,q}}{\overset{\mathcal{D}_{1,q}}{\gtrless}} 1
\Rightarrow
\|\boldsymbol{y}_{q}\|^2
\underset{\mathcal{D}_{0,q}}{\overset{\mathcal{D}_{1,q}}{\gtrless}}
N\frac{\Lambda_{0,q}\Lambda_{1,q}}{\Lambda_{1,q}-\Lambda_{0,q}}
\ln\frac{\Lambda_{1,q}}{\Lambda_{0,q}}
\triangleq \vartheta_q^{*}.
\end{equation}
Accordingly, the optimal detection rule in \eqref{eq:LRT} can be implemented using an energy detector.
Under hypothesis $\mathcal{H}_i$, $\|\boldsymbol{y}_{q}\|^2$ is the sum of $N$ independent exponential random variables, each with mean $\Lambda_{i,q}$, and thus follows a Gamma distribution. The two error probabilities are obtained from its cumulative distribution function, and the minimum DEP is given by
\begin{equation}\label{eq:min_DEP}
\begin{aligned}
\xi_q^*
&= \mathbb{P}\!\left\{\|\boldsymbol{y}_{q}\|^2 \le \vartheta_q^{*}\big|\mathcal{H}_1\right\}
+ \mathbb{P}\!\left\{\|\boldsymbol{y}_{q}\|^2 \ge \vartheta_q^{*}\big|\mathcal{H}_0\right\}
\\
&= 1 - \underbrace{\left[
\frac{\gamma_{\rm inc}\!\left(N,\vartheta_q^{*}/\Lambda_{0,q}\right)}{\Gamma(N)}
-
\frac{\gamma_{\rm inc}\!\left(N,\vartheta_q^{*}/\Lambda_{1,q}\right)}{\Gamma(N)}
\right]}_{
=\,\mathcal{V}_{T}(\mathbb{P}\left(\boldsymbol{y}_{q}|\mathcal{H}_0\right),\mathbb{P}\left(\boldsymbol{y}_{q}|\mathcal{H}_1\right))},
\end{aligned}
\end{equation}
where $\gamma_{\rm inc}(N,x) \triangleq \int_0^x e^{-t}t^{N-1}\mathrm{d}t$ is the lower incomplete Gamma function, $\Gamma(N) = (N-1)!$ is the Gamma function, and $\mathcal{V}_{T}(\mathbb{P}\left(\boldsymbol{y}_{q}|\mathcal{H}_0\right),\mathbb{P}\left(\boldsymbol{y}_{q}|\mathcal{H}_1\right))$ is the total variation distance between the PDFs $\mathbb{P}\left(\boldsymbol{y}_{q}|\mathcal{H}_0\right)$ and $\mathbb{P}\left(\boldsymbol{y}_{q}|\mathcal{H}_1\right)$.

Based on the detection performance in \eqref{eq:min_DEP}, the covertness requirement at warden $q$ is expressed as
\begin{equation}\label{eq:DEP_constraint}
\xi_q^* = 1 - \mathcal{V}_{T}(\mathbb{P}\left(\boldsymbol{y}_{q}|\mathcal{H}_0\right),\mathbb{P}\left(\boldsymbol{y}_{q}|\mathcal{H}_1\right))\ge 1 - \epsilon, \forall q \in \mathcal{Q},
\end{equation}
where $\epsilon \in [0, 1]$ is the given covertness level, and a smaller $\epsilon$ means a stricter covertness requirement. In particular, when $\epsilon = 0$, we have $\xi_q^* = 1$, in which case warden $q$ cannot do better than a random guess.
The expression for $\xi_q^*$ is difficult to handle directly because it involves the lower incomplete Gamma functions in \eqref{eq:min_DEP}.
Although the exact DEP constraint can be expressed using a numerically computed power-ratio threshold, we follow~\cite{Yan2019Gaussian} and adopt the KL-based sufficient condition for its simple analytical form and explicit covertness guarantee.
Specifically, we use a lower bound on $\xi_q^*$ based on the Kullback--Leibler (KL) divergence, given by
\begin{equation}\label{eq:KL_div}
\begin{aligned}
&\mathcal{D}(\mathbb{P}\left(\boldsymbol{y}_{q}|\mathcal{H}_1\right)\!\|\mathbb{P}\left(\boldsymbol{y}_{q}|\mathcal{H}_0\right)) \!=\! \int \!\mathbb{P}\left(\boldsymbol{y}_{q}|\mathcal{H}_1\right)
\ln\!\left(\frac{\mathbb{P}\left(\boldsymbol{y}_{q}|\mathcal{H}_1\right)}{\mathbb{P}\left(\boldsymbol{y}_{q}|\mathcal{H}_0\right)}\right)\mathrm{d}\boldsymbol{y}_{q}\\
&=\! N\!\left[\!\frac{\Lambda_{1,q}(\boldsymbol{x})}{\Lambda_{0,q}(\boldsymbol{x})} \!-\! 1 \!-\! \ln\!\frac{\Lambda_{1,q}(\boldsymbol{x})}{\Lambda_{0,q}(\boldsymbol{x})}\!\right]\!=\! N\!f\!\left(\!\frac{\Lambda_{1,q}(\boldsymbol{x}) - \Lambda_{0,q}(\boldsymbol{x})}{\Lambda_{0,q}(\boldsymbol{x})}\!\right)\!,
\end{aligned}
\end{equation}
where $f(x) \triangleq x - \ln(1+x)$ is monotonically increasing for $x \ge 0$.
According to \emph{Pinsker's inequality}, we have
\begin{equation}\label{eq:pinsker}
\mathcal{V}_{T}(\mathbb{P}\left(\boldsymbol{y}_{q}|\mathcal{H}_0\right),\mathbb{P}\left(\boldsymbol{y}_{q}|\mathcal{H}_1\right))\le \sqrt{\frac{\mathcal{D}(\mathbb{P}\left(\boldsymbol{y}_{q}|\mathcal{H}_1\right)\|\mathbb{P}\left(\boldsymbol{y}_{q}|\mathcal{H}_0\right))}{2}}.
\end{equation}
Therefore, \eqref{eq:DEP_constraint} holds as long as $\mathcal{D}(\mathbb{P}\left(\boldsymbol{y}_{q}|\mathcal{H}_1\right)\|\mathbb{P}\left(\boldsymbol{y}_{q}|\mathcal{H}_0\right)) \le 2\epsilon^2$.
By substituting the expressions of $\Lambda_{1,q}(\boldsymbol{x})$ and $\Lambda_{0,q}(\boldsymbol{x})$ into the argument of $f$, we have
\begin{equation}
\frac{\Lambda_{1,q}(\boldsymbol{x}) - \Lambda_{0,q}(\boldsymbol{x})}{\Lambda_{0,q}(\boldsymbol{x})}
= \frac{\boldsymbol{h}_{q}^H(\boldsymbol{x}) \!\left(\sum_{k\in\mathcal{K}_c}\boldsymbol{w}_k \boldsymbol{w}_k^H\right)\! \boldsymbol{h}_{q}(\boldsymbol{x})}
{\Lambda_{0,q}(\boldsymbol{x})}.
\end{equation}
Based on the monotonicity of $f(x)$, the KL divergence condition
\(
\mathcal{D}(\mathbb{P}\left(\boldsymbol{y}_{q}|\mathcal{H}_1\right)\|\mathbb{P}\left(\boldsymbol{y}_{q}|\mathcal{H}_0\right)) \le 2\epsilon^2
\)
can be equivalently expressed as $(\Lambda_{1,q}(\boldsymbol{x})-\Lambda_{0,q}(\boldsymbol{x}))/\Lambda_{0,q}(\boldsymbol{x}) \le \eta$, where $\eta \ge 0$ is the unique solution of $f(\eta) = 2\epsilon^2/N$.
Rearranging the resulting inequality, we obtain the following equivalent quadratic form of the KL divergence condition at each warden, given by
\begin{equation}\label{eq:covert_compact}
\boldsymbol{h}_{q}^H(\boldsymbol{x}) \, \bm\Xi \, \boldsymbol{h}_{q}(\boldsymbol{x}) \le \eta \, \sigma_{q}^2, \forall q \in \mathcal{Q},
\end{equation}
where
$\bm\Xi\!\triangleq\!\sum_{k\in\mathcal{K}_c}\!\!\boldsymbol{w}_k \boldsymbol{w}_k^H\!-\!\eta\!\left(\sum_{k\in\mathcal{K}_o}\!\!\boldsymbol{w}_k\boldsymbol{w}_k^H \!+\! \boldsymbol{R}_{\rm A}\right)\!\in\!\mathbb{C}^{M\times M}.$


\subsection{Imperfect CSI Models}
Due to the absence of pilot transmissions from warden $q$, Alice cannot estimate $\boldsymbol{h}_{q}(\boldsymbol{x})$ from reference signals~\cite{ForouzeshCDI}. In practice, the AoD $\theta_q$ is unlikely to be perfectly known, and the channel gain $\beta_q$ is also subject to estimation errors. Both factors lead to imperfect warden CSI in \eqref{eq:warden_channel}. Accordingly, the warden channel is expressed as the sum of its estimated value and estimation error, i.e.,
\begin{equation}\label{eq:CSI_error_model}
\boldsymbol{h}_{q}(\boldsymbol{x}) = \bar{\boldsymbol h}_q(\boldsymbol{x}) + \Delta\boldsymbol h_q, \forall q\in\mathcal{Q},
\end{equation}
where $\bar{\boldsymbol h}_q(\boldsymbol{x})$ denotes the nominal warden channel determined by the estimated AoD and channel gain, and $\Delta\boldsymbol h_q$ is the corresponding CSI error.
In the following, we introduce two CSI error models.

\begin{itemize}
\item \emph{Bounded CSI error}: The squared norm of the CSI error is bounded by the known constant $e_q^2$, i.e.,
\begin{equation}\label{eq:bounded_CSI_error}
\Delta\boldsymbol h_q \in \mathcal{W}_q, \forall q\in\mathcal Q,
\end{equation}
where $\mathcal W_q \triangleq \{\Delta\boldsymbol h_q | \|\Delta\boldsymbol h_q\|^2 \le e_q^2\}$.
\item \emph{Statistical CSI error}: The CSI error vector is characterized by a zero-mean complex Gaussian distribution with a known covariance matrix, i.e.,
\begin{equation}
\Delta\boldsymbol h_q \sim \mathcal{CN}(\bm{0}, \boldsymbol{E}_{q}), \forall q\in\mathcal Q,
\end{equation}
where $\boldsymbol{E}_{q}\succeq \boldsymbol{0}$ denotes the CSI error covariance matrix known at Alice.
\end{itemize}

\section{Worst-case Robust Beamforming}\label{sec:wc_csi_design}
In this section, we first formulate the robust optimization problem under the bounded CSI error model and develop a BCD-based algorithm incorporating bisection search, SDR, and SCA methods to solve the intractable problem.
\subsection{Problem Formulation}
We aim to maximize the minimum weighted SINR among all legitimate users subject to the covertness constraints at all wardens, the transmit power budget, and the MA position constraints. Specifically, the optimization problem can be given by
\begin{subequations}\label{p1}
\begin{align}
\mathmakebox[2em][r]{\underset{\mathclap{\tau,\boldsymbol{w}_k,\boldsymbol{R}_{\rm A},\boldsymbol{x}}}{\max}}\; & \quad\tau \\
\mathmakebox[2em][r]{\text{s.t.}}\;
&\quad \gamma_{1,k}^C(\boldsymbol{w}_k,\boldsymbol{R}_{\rm A},\boldsymbol{x}) \geq \nu_k\tau,\forall k\in\mathcal{K} \label{Const_1}\\
&\quad \boldsymbol{h}_{q}^H(\boldsymbol{x}) \, \bm\Xi \, \boldsymbol{h}_{q}(\boldsymbol{x}) \le \eta \, \sigma_{q}^2,\Delta\boldsymbol h_q \in \mathcal{W}_q, \forall q\in\mathcal{Q},\label{Const_2}\\
&\quad \sum_{k\in\mathcal{K}}(\boldsymbol{w}_k^H\boldsymbol{w}_k) + \mathrm{Tr}(\boldsymbol{R}_{\rm A}) \leq P, \label{Const_3}\\
&\quad x_{m+1}-x_m \geq x_{th}, \forall m\in\mathcal{M}\setminus\{M\}, \label{Const_4}\\
&\quad x_m\in \mathcal{X},\forall m\in\mathcal{M}, \label{Const_5}
\end{align}
\end{subequations}
where $\nu_k>0$ in \eqref{Const_1} is the priority weight of user $k$, $P$ in \eqref{Const_3} denotes Alice's transmit power budget, and \eqref{Const_4} enforces a minimum spacing of $x_{th} \ge 0$ between transmit MAs to avoid coupling.
Since overt transmission also provides cover for covert transmission, we set $\nu_k=1$ for overt users $k\in\mathcal K_o$ and $\nu_k=\nu$ for covert users $k\in\mathcal K_c$, where $\nu>0$ controls the priority of covert users relative to overt users.
For a given $\tau$, a larger $\nu$ requires a higher SINR for covert users.

The formulated problem is difficult to solve directly due to the coupling among the beamforming, AN covariance matrix, and MA positions in the quality-of-service (QoS) and covertness constraints.
To this end, we propose a BCD-based algorithm by alternately optimizing the beamforming and AN covariance matrix via semidefinite relaxation and a bisection search over the SINR threshold, and updating MA positions via the SCA technique.
\subsection{Optimizing Beamforming and AN Covariance Matrix}
With the MA positions being fixed, we focus on optimizing the transmit beamforming vectors $\boldsymbol{w}_k$ and the AN covariance matrix $\boldsymbol{R}_{\rm A}$. The non-convexity of problem \eqref{p1} lies in the quadratic terms with respect to (w.r.t.) $\boldsymbol{w}_k$ in \eqref{Const_1} and \eqref{Const_2}. 
To deal with this issue, we employ the SDR method~\cite{wtl_noma_isac}. 
Towards this end, we define $\boldsymbol{W}_k \triangleq \boldsymbol{w}_k\boldsymbol{w}_k^H$ and impose $\boldsymbol{W}_k\succeq\boldsymbol0$ and $\mathrm{rank}(\boldsymbol{W}_k)=1$, $\forall k\in\mathcal K$. Then, \eqref{Const_1} can be reformulated as
\begin{equation}
\frac{\boldsymbol{h}_k^H(\boldsymbol{x}) \boldsymbol{W}_k \boldsymbol{h}_k(\boldsymbol{x})}
{\boldsymbol{h}_k^H(\boldsymbol{x})(\boldsymbol{R} - \boldsymbol{W}_k)\boldsymbol{h}_k(\boldsymbol{x}) + \sigma_k^2} \ge \nu_k\tau,
\end{equation}
where $\boldsymbol{R} = \sum_{k\in\mathcal{K}} \boldsymbol{W}_k + \boldsymbol{R}_{\rm A}$, and $\nu_k$ remain constants on the right-hand side (RHS).

To tackle semi-infinite constraints introduced by channel uncertainty in \eqref{Const_2}, we introduce the following lemma.
\begin{lemma}[S-Procedure]\phantomsection\label{lem:S_procedure}
\it Define the quadratic function $f_m(\bm x)$ w.r.t.\ $\bm x\in\mathbb{C}^{M\times 1}$ as
\begin{equation*}
f_m(\bm x) = \bm x^H \mathbf{A}_m \bm x + 2\,\Re\{\mathbf{b}_m^H \bm x\} + c_m,\; m=1,2,
\end{equation*}
where $\mathbf{A}_m \in \mathbb{C}^{M\times M}$, $\mathbf{b}_m \in \mathbb{C}^{M\times 1}$, and $c_m \in \mathbb{R}$.
The condition $f_1(\bm x) \le 0 \Rightarrow f_2(\bm x)\le 0$ holds if and only if there exists $\omega \ge 0$ such that
\begin{equation}\label{eq:S_proc_LMI}
\omega\!\begin{bmatrix} \mathbf{A}_1 & \mathbf{b}_1\\ \mathbf{b}_1^H & c_1\end{bmatrix}
- \begin{bmatrix} \mathbf{A}_2 & \mathbf{b}_2\\ \mathbf{b}_2^H & c_2\end{bmatrix}
\succeq \mathbf{0}_{M+1}.
\end{equation}
\end{lemma}
By substituting \eqref{eq:CSI_error_model} into (\ref{Const_2}), we have 
\begin{equation}\label{eq:covert_constraint_bounded_expanded}
\begin{aligned}
\Delta\boldsymbol h_q^H \bm\Xi \Delta\boldsymbol h_q +& 2\Re\{\bar{\boldsymbol h}_q^H \bm\Xi \Delta\boldsymbol h_q\} \\
&\quad+ \bar{\boldsymbol h}_q^H \bm\Xi \bar{\boldsymbol h}_q - \eta \sigma_q^2 \le 0, \forall q\in\mathcal{Q}.
\end{aligned}
\end{equation}
Note that \eqref{eq:bounded_CSI_error} leads to $\Delta\boldsymbol h_q^H \Delta\boldsymbol h_q \le e_q^2$, $\forall q\in\mathcal Q$.
Lemma~\ref{lem:S_procedure} is employed to tackle the CSI uncertainty in the covertness constraint in~\eqref{eq:covert_constraint_bounded_expanded}. 
Specifically, the constraint corresponding to each warden in \eqref{Const_2} can be recast as follows
\begin{equation}
\begin{aligned}
&\mathbf{A}_1 = \mathbf{I}_M, \mathbf{b}_1 = \mathbf{0}, c_1 = -e_q^2,\\ & \mathbf{A}_2 = \bm\Xi, \mathbf{b}_2 = \bm\Xi\bar{\boldsymbol h}_q(\boldsymbol x),
c_2 = \bar{\boldsymbol h}_q^H(\boldsymbol x)\bm\Xi\bar{\boldsymbol h}_q(\boldsymbol x) - \eta\sigma_{q}^2.
\end{aligned}
\end{equation}
Then, \eqref{Const_2} is equivalently expressed as the following LMI
\begin{equation}\label{eq:Sproc_lmi}
\begin{aligned}
&\bm\Theta_q(\boldsymbol x,\{\boldsymbol W_k\},\boldsymbol R,\omega_q)
\triangleq \\&
\begin{bmatrix}
\omega_q\mathbf{I}_M-\bm\Xi
&-\bm\Xi\bar{\boldsymbol h}_q(\boldsymbol x)\\
-\bar{\boldsymbol h}_q^H(\boldsymbol x)\bm\Xi
&\eta\sigma_{q}^2-
\bar{\boldsymbol h}_q^H(\boldsymbol x)\bm\Xi\bar{\boldsymbol h}_q(\boldsymbol x)-\omega_q e_{q}^2
\end{bmatrix}
\succeq\boldsymbol0,
\end{aligned}
\end{equation}
where $\boldsymbol{\omega}=[\omega_1,\ldots,\omega_Q]^{\mathrm T}\ge\boldsymbol0$ is the vector of slack variables and $\bm\Xi=(1+\eta)\sum_{k\in\mathcal{K}_c}\boldsymbol{W}_k-\eta\boldsymbol{R}$.

Then, for any given $\tau\ge0$, the beamforming and AN covariance matrix subproblem in~\eqref{p1} can be reformulated as the following feasibility problem, i.e.,
\begin{subequations}\label{p2}
\begin{align}
\mathmakebox[0pt][r]{\text{find}}\; & \quad\boldsymbol{W}_k,\boldsymbol{R},\boldsymbol{\omega}\\
\mathmakebox[0pt][r]{\text{s.t.}}\;
&\quad\frac{\boldsymbol{h}_k^H(\boldsymbol{x}) \boldsymbol{W}_k \boldsymbol{h}_k(\boldsymbol{x})}
{\boldsymbol{h}_k^H(\boldsymbol{x}) (\boldsymbol{R} \!-\!  \boldsymbol{W}_k) \boldsymbol{h}_k(\boldsymbol{x}) \!+\! \sigma_k^2} \geq \nu_k\tau, \forall k\in\mathcal{K},\label{Const_p2_sinr}\\
&\quad\bm\Theta_q(\boldsymbol x,\boldsymbol W_k,\boldsymbol R,\boldsymbol\omega) \succeq \boldsymbol0,\boldsymbol\omega\ge 0,\label{Const_p2_covert_lmi}\\
&\quad\mathrm{Tr}(\boldsymbol{R}) \le P,\label{Const_p2_power}\\
&\quad\boldsymbol{R}-\sum_{k\in\mathcal{K}}\boldsymbol{W}_k\succeq \boldsymbol{0},\boldsymbol{W}_k \succeq \boldsymbol{0}, \forall k\in\mathcal{K},\label{Const_p2_split}\\
&\quad\mathrm{rank}(\boldsymbol{W}_k) = 1, \forall k\in\mathcal{K}.\label{Const_p2_rank}
\end{align}
\end{subequations}
Omitting the non-convex rank-one constraints \eqref{Const_p2_rank}, we obtain a relaxed version of the above problem.
Thus, the beamforming and AN covariance matrix optimization reduces to a series of semidefinite programming (SDP) feasibility subproblems, which can be solved using the CVX toolbox~\cite{GrantCVX2014}.

Note that for any fixed $\tau\ge0$, the relaxed problem~\eqref{p2} is an SDP.
Therefore, the corresponding relaxed weighted max-min SINR problem is quasi-concave and can be solved by a bisection search over $\tau$.
The details of the bisection search method are summarized in Algorithm~\ref{alg:robust_bisection}.

Let $\{\overline{\boldsymbol{W}}_k\}_{k=1}^{K}$ and $\overline{\boldsymbol{R}}$ denote a feasible solution of the relaxed SDP for a given $\tau>0$.
Here, we note that if $\mathrm{rank}(\overline{\boldsymbol{W}}_k)=1$ holds for all $k\in\mathcal K$, the solution to the relaxed problem is also feasible for the original rank-constrained problem~\eqref{p2}.
Otherwise, we can reconstruct a rank-one feasible solution in closed form that attains the same SINR threshold.
Specifically, the beamforming vectors $\boldsymbol{w}_k^\star$, the rank-one matrices $\boldsymbol{W}_k^\star$, and the corresponding AN covariance matrix $\boldsymbol{R}_{\rm A}^\star$ can be reconstructed as follows
\begin{equation}\label{SDR_optimal}
\begin{aligned}
&\boldsymbol{R}^\star = \overline{\boldsymbol{R}},\boldsymbol{w}_k^\star = (\boldsymbol{h}_k^H(\boldsymbol{x})\overline{\boldsymbol{W}}_k\boldsymbol{h}_k(\boldsymbol{x}))^{-1/2}\overline{\boldsymbol{W}}_k\boldsymbol{h}_k(\boldsymbol{x}), \\
  &\boldsymbol{W}_k^\star = \boldsymbol{w}_k^\star\boldsymbol{w}_k^{\star H}, \boldsymbol{R}_{\rm A}^\star = \boldsymbol{R}^\star - \sum_{k=1}^{K}\boldsymbol{W}_k^\star.
\end{aligned}
\end{equation}

The proof is provided in Appendix \ref{AppSDR}.

\subsection{Optimizing Auxiliary Variables}\label{subsec:auxiliary_update}
With other variables being fixed, we focus on the optimization of the auxiliary variables.
For each user, we introduce an auxiliary variable $\alpha_k\in\mathbb C$ and apply the quadratic transform as follows
\begin{equation}\label{eq:qt_position_sinr}
\begin{aligned}
\tilde\gamma_{1,k}^C(\boldsymbol x,\alpha_k)
=&2\Re\{\alpha_k^*\boldsymbol h_k^H(\boldsymbol x)\boldsymbol w_k\}\\
&-|\alpha_k|^2\left(\boldsymbol h_k^H(\boldsymbol x)\boldsymbol\Xi_k\boldsymbol h_k(\boldsymbol x)+\sigma_k^2\right),
\end{aligned}
\end{equation}
where $\boldsymbol\Xi_k\triangleq \sum_{i\in\mathcal K\setminus\{k\}}\boldsymbol W_i+\boldsymbol{R}_{\rm A}$.
$\tilde\gamma_{1,k}^C(\boldsymbol x,\alpha_k)$ is concave w.r.t.\ $\alpha_k$, and its optimal auxiliary variable has the closed-form solution
\begin{equation}\label{eq:alpha_update_position}
\alpha_k^\star
=\frac{\boldsymbol h_k^H(\boldsymbol x)\boldsymbol w_k}
{\boldsymbol h_k^H(\boldsymbol x)\boldsymbol\Xi_k\boldsymbol h_k(\boldsymbol x)+\sigma_k^2},
\quad\forall k\in\mathcal K.
\end{equation}

\subsection{Optimizing Transmit Antenna Positions}\label{subsec:MA_position_update_wc}
With $\boldsymbol W_k$, $\boldsymbol{R}_{\rm A}$, and $\alpha_k$ fixed, we jointly optimize the MA positions $\boldsymbol x$ and the SINR threshold $\tau$.
Accordingly, the MA position subproblem can be formulated as
\begin{subequations}\label{p3_raw}
\begin{align}
\mathmakebox[2em][r]{\underset{\mathclap{\boldsymbol x,\tau}}{\max}}\;
&\quad\tau
\\
\mathmakebox[2em][r]{\text{s.t.}}\;
&\quad\gamma_{1,k}^C(\boldsymbol x)\ge \nu_k\tau,\quad \forall k\in\mathcal K,
\label{Const_p3_raw_sinr}\\
&\quad\left(\bar{\boldsymbol h}_q(\boldsymbol x)+\bm\delta_q\right)^H
\bm\Xi
\left(\bar{\boldsymbol h}_q(\boldsymbol x)+\bm\delta_q\right)
\le \eta\sigma_q^2, \notag\\
&\quad\bm\delta_q^H\bm\delta_q \le e_q^2,\ \forall q\in\mathcal Q,
\label{Const_p3_raw_covert}\\
&\quad(\ref{Const_4}),(\ref{Const_5})
\end{align}
\end{subequations}
where $\bm\Xi
=(1+\eta)\sum_{k\in\mathcal K_c}\boldsymbol W_k
-\eta\boldsymbol{R}$ and $\bm\delta_q$ denotes the channel estimation error $\Delta\boldsymbol h_q$ under the bounded CSI error model in \eqref{eq:bounded_CSI_error}.
It is worth noting that the non-convexity of problem~\eqref{p3_raw} lies in the SINR constraint \eqref{Const_p3_raw_sinr} and the covertness constraint \eqref{Const_p3_raw_covert}.
To tackle this challenge, we employ SCA to construct a sequence of convex subproblems.
 
To approximate the two terms involving $\boldsymbol x$ in \eqref{eq:qt_position_sinr} separately, we define $a_k(\boldsymbol x)$ and $b_k(\boldsymbol x)$ as
\begin{equation}\label{eq:ak_position_def}
a_k(\boldsymbol x)
\triangleq
2\Re\{\alpha_k^*\boldsymbol h_k^H(\boldsymbol x)\boldsymbol w_k\},
\end{equation}
\begin{equation}\label{eq:bk_position_def}
b_k(\boldsymbol x)
\triangleq
|\alpha_k|^2\boldsymbol h_k^H(\boldsymbol x)\boldsymbol\Xi_k\boldsymbol h_k(\boldsymbol x).
\end{equation}
To construct a concave surrogate function for the SINR in \eqref{eq:qt_position_sinr}, we introduce the following lemma.
\begin{lemma}[Global Lower Bound Approximation]\phantomsection\label{lem:global_lower_bound}
\it Let $f(\boldsymbol x):\mathbb R^N\to\mathbb R$ denote a continuously differentiable function with bounded curvature, i.e., there exists a positive real number $\mu$ such that $\mu\ge\|\nabla^2 f(\boldsymbol x)\|_2$ for any $\boldsymbol x$.
Define $f_2(\boldsymbol x)=-f(\boldsymbol x)$.
Then, for all $\boldsymbol x,\boldsymbol y\in\mathbb R^N$, the following inequality holds
\begin{equation}\label{eq:global_lower_bound}
f_2(\boldsymbol x)
\ge{} f_2(\boldsymbol y)+\nabla f_2(\boldsymbol y)^T(\boldsymbol x-\boldsymbol y)
-\frac{1}{2}\mu(\boldsymbol x-\boldsymbol y)^T(\boldsymbol x-\boldsymbol y).
\end{equation}
\end{lemma}
\textit{Proof:} See Appendix~\ref{AppGlobalLowerBound} for the proof.

Based on Lemma~\ref{lem:global_lower_bound}, we approximate $a_k(\boldsymbol x)$ and $b_k(\boldsymbol x)$ at $\boldsymbol x^{(t)}$ by constructing a concave lower bound and a convex upper bound, respectively.
Specifically, by applying the lemma with $f_2=a_k$ and $f_2=-b_k$, we can obtain
\begin{equation}\label{eq:ak_under_position}
\begin{aligned}
a_k(\boldsymbol x)
\ge{} a_k(\boldsymbol x^{(t)})
&+\nabla a_k(\boldsymbol x^{(t)})^T(\boldsymbol x-\boldsymbol x^{(t)})
\\
&-\frac{\varpi_{a,k}}{2}\|\boldsymbol x-\boldsymbol x^{(t)}\|_2^2
\triangleq
\underline a_k(\boldsymbol x),
\end{aligned}
\end{equation}
\begin{equation}\label{eq:bk_over_position}
\begin{aligned}
b_k(\boldsymbol x)
\le{} b_k(\boldsymbol x^{(t)})
&+\nabla b_k(\boldsymbol x^{(t)})^T(\boldsymbol x-\boldsymbol x^{(t)})
\\
&+\frac{\varpi_{b,k}}{2}\|\boldsymbol x-\boldsymbol x^{(t)}\|_2^2
\triangleq \bar b_k(\boldsymbol x),
\end{aligned}
\end{equation}
where $\nabla a_k(\boldsymbol x)$ and $\nabla b_k(\boldsymbol x)$ denote the gradient vectors of $a_k(\boldsymbol x)$ and $b_k(\boldsymbol x)$ w.r.t.\ $\boldsymbol x$, respectively.
The nonnegative constants $\varpi_{a,k}$ and $\varpi_{b,k}$ satisfy $\varpi_{a,k}\ge\|\nabla^2a_k(\boldsymbol x)\|_2$ and $\varpi_{b,k}\ge\|\nabla^2b_k(\boldsymbol x)\|_2$ for all $\boldsymbol x$, where $\nabla^2a_k(\boldsymbol x)$ and $\nabla^2b_k(\boldsymbol x)$ are the corresponding Hessian matrices.
At $\boldsymbol x^{(t)}$, $\underline a_k$ and $\bar b_k$ match the values and gradients of $a_k$ and $b_k$, respectively, and the curvature bounds ensure their global validity.
The expressions of $\varpi_{a,k}$ and $\varpi_{b,k}$ are derived in Appendix~\ref{AppSINR_a}.
Combining \eqref{eq:ak_under_position} and \eqref{eq:bk_over_position}, we obtain the concave SINR lower bound as
$
\widehat\gamma_{1,k}^C(\boldsymbol x)
\triangleq
\underline a_k(\boldsymbol x)
-
\bar b_k(\boldsymbol x)
-|\alpha_k|^2\sigma_k^2
\le
\tilde\gamma_{1,k}^C(\boldsymbol x,\alpha_k).
$
The resulting conservative convex constraint is
\begin{equation}\label{eq:sinr_sca_constraint_position}
\widehat\gamma_{1,k}^C(\boldsymbol x)\ge\nu_k\tau,
\quad \forall k\in\mathcal K.
\end{equation}

Next, we construct a convex approximation of the covertness constraint~\eqref{Const_2}. By substituting $\boldsymbol h_q(\boldsymbol x)=\bar{\boldsymbol h}_q(\boldsymbol x)+\bm\delta_q$ into the left-hand side (LHS) of \eqref{Const_2}, we can obtain the equivalent form
\begin{equation}\label{eq:covert_constraint_expanded_position}
\bm\delta_q^H\bm\Xi\bm\delta_q
+
\underbrace{\bar{\boldsymbol h}_q^H(\boldsymbol x)\bm\Xi\bm\delta_q
+
\bm\delta_q^H\bm\Xi\bar{\boldsymbol h}_q(\boldsymbol x)}_{\triangleq\mu_q(\boldsymbol x,\bm\delta_q)}
+
\underbrace{\bar{\boldsymbol h}_q^H(\boldsymbol x)\bm\Xi\bar{\boldsymbol h}_q(\boldsymbol x)}_{\triangleq\varkappa_q(\boldsymbol x)}
\le \eta\sigma_q^2.
\end{equation}
Using the uniform Hessian bound $\varpi_{\mu,q}$, we construct the following convex quadratic upper bound on $\mu_q(\boldsymbol x,\bm\delta_q)$
\begin{equation}\label{eq:mu_upper_two_part_app}
\begin{aligned}
\mu_q(\boldsymbol{x},\bm\delta_q)
\le
\mu_q&(\boldsymbol{x}^{(t)},\bm\delta_q)
+ \nabla_{\boldsymbol{x}}\mu_q(\boldsymbol{x}^{(t)},\bm\delta_q)^T
(\boldsymbol{x}-\boldsymbol{x}^{(t)})
\\
& + \frac{\varpi_{\mu,q}}{2}
\|\boldsymbol{x}-\boldsymbol{x}^{(t)}\|_2^2 \triangleq \bar{\mu}_q^{(t)}(\boldsymbol{x},\bm\delta_q),
\end{aligned}
\end{equation}
where $\varpi_{\mu,q}$ is given in Appendix~\ref{AppCovertSCA}.
Similarly, $\varkappa_q(\boldsymbol x)$ can be upper-bounded as follows
\begin{equation}\label{eq:varkappa_q_upper_bound}
\begin{aligned}
\varkappa_q(\boldsymbol x)\le
\varkappa_q(\boldsymbol{x}^{(t)})
&+ \nabla_{\boldsymbol{x}}\varkappa_q(\boldsymbol{x}^{(t)})^T
(\boldsymbol{x}-\boldsymbol{x}^{(t)})
\\&
+ \frac{\varpi_{\varkappa,q}}{2}
\|\boldsymbol{x}-\boldsymbol{x}^{(t)}\|_2^2 \triangleq \bar{\varkappa}_q^{(t)}(\boldsymbol{x}),
\end{aligned}
\end{equation}
where $\varpi_{\varkappa,q}$ is given in Appendix~\ref{AppSINR_a}.
Then, substituting \eqref{eq:mu_upper_two_part_app} and \eqref{eq:varkappa_q_upper_bound} into the LHS of \eqref{eq:covert_constraint_expanded_position}, we have
\begin{equation}\label{eq:covert_sca_expanded_position}
\bm\delta_q^H\bm\Xi\bm\delta_q
+
\bar{\mu}_q^{(t)}(\boldsymbol{x},\bm\delta_q)
+\bar{\varkappa}_q^{(t)}(\boldsymbol{x})
-\eta\sigma_q^2
\le 0.
\end{equation}
We rewrite $\nabla_{\boldsymbol{x}}\mu_q(\boldsymbol{x}^{(t)},\bm\delta_q)^T(\boldsymbol{x}-\boldsymbol{x}^{(t)})$ as
\begin{equation}\label{eq:mu_linear_compact_app}
\nabla_{\boldsymbol{x}}\mu_q(\boldsymbol{x}^{(t)},\bm\delta_q)^T
(\boldsymbol{x}-\boldsymbol{x}^{(t)}) \!=\!
\boldsymbol z_q^{(t)}(\boldsymbol{x})^H\bm\delta_q
\!+\!
\bm\delta_q^H\boldsymbol z_q^{(t)}(\boldsymbol{x}),
\end{equation}
where
$ \boldsymbol z_q^{(t)}(\boldsymbol{x})
\triangleq
\left[
\boldsymbol r_{q,1}^H(x_1^{(t)}),\,
\ldots,\,
\boldsymbol r_{q,M}^H(x_M^{(t)})
\right]
(\boldsymbol{x}-\boldsymbol{x}^{(t)}).$
Hence, \eqref{eq:covert_sca_expanded_position} can be written in the following tractable form
\begin{equation}\label{eq:covert_sca_explicit_app}
\bm\delta_q^H\bm\Xi\bm\delta_q
\!+\!
\bm\delta_q^H
\boldsymbol c_q^{(t)}(\boldsymbol{x})
\!+\!
\boldsymbol c_q^{(t)}(\boldsymbol{x})^H\bm\delta_q
\!+\! \upsilon_q^{(t)}(\boldsymbol{x}) \!-\! \eta\sigma_q^2
\le 0,
\end{equation}
where $\boldsymbol c_q^{(t)}(\boldsymbol{x}) \triangleq \bm\Xi\bar{\boldsymbol h}_q(\boldsymbol{x}^{(t)}) + \boldsymbol z_q^{(t)}(\boldsymbol{x})$, and
$\upsilon_q^{(t)}(\boldsymbol{x})
= \varkappa_q(\boldsymbol{x}^{(t)})
+ \nabla_{\boldsymbol{x}}\varkappa_q(\boldsymbol{x}^{(t)})^T
(\boldsymbol{x}-\boldsymbol{x}^{(t)})
+\frac{\varpi_{\varkappa,q}+\varpi_{\mu,q}}{2}\|\boldsymbol{x}-\boldsymbol{x}^{(t)}\|_2^2$.
The upper bounds in \eqref{eq:mu_upper_two_part_app} and \eqref{eq:varkappa_q_upper_bound} are globally valid and match the function values and position gradients at $\boldsymbol x^{(t)}$. Thus, \eqref{eq:covert_sca_explicit_app} implies \eqref{eq:covert_constraint_expanded_position} for every $\|\bm\delta_q\|_2\le e_q$ and retains the current feasible position.
By the S-procedure, \eqref{eq:covert_sca_explicit_app} is equivalent to the existence of $\omega_q\ge0$ such that
\begin{equation}\label{eq:position_covert_sca_lmi_main}
\begin{bmatrix}
\omega_q\mathbf{I}_M-\bm\Xi
&
-\boldsymbol c_q^{(t)}(\boldsymbol x)\\
-\boldsymbol c_q^{(t)}(\boldsymbol x)^H
&
\eta\sigma_q^2-\upsilon_q^{(t)}(\boldsymbol x)-\omega_q e_q^2
\end{bmatrix}
\succeq\boldsymbol0.
\end{equation}
The matrix inequality in \eqref{eq:position_covert_sca_lmi_main} is still nonlinear w.r.t.\ $\boldsymbol x$ due to the quadratic term in $\upsilon_q^{(t)}(\boldsymbol x)$.
To obtain an LMI formulation, we introduce the following lemma.
\begin{lemma}\phantomsection\label{lem:lmi_brc_relaxation}
\it For any complex Hermitian matrix $\boldsymbol A\in\mathbb{C}^{N\times N}$, $\boldsymbol b\in\mathbb{C}^{N\times 1}$, $c\in\mathbb R$, and $d\in\mathbb R$, if
\begin{equation}
\begin{bmatrix}
\boldsymbol A & \boldsymbol b\\
\boldsymbol b^H & c
\end{bmatrix}\succeq\boldsymbol 0
\quad\text{and}\quad d\ge c,
\end{equation}
then
\begin{equation}\label{eq:brc_lmi_impl}
\begin{bmatrix}
\boldsymbol A & \boldsymbol b\\
\boldsymbol b^H & d
\end{bmatrix}\succeq\boldsymbol 0
\end{equation}
\end{lemma}

By applying Lemma~\ref{lem:lmi_brc_relaxation}, the matrix inequality \eqref{eq:position_covert_sca_lmi_main} is equivalently reformulated as the following LMI together with the auxiliary constraint \eqref{eq:48}
\begin{equation}\label{eq:position_covert_sca_lmi_relaxed}
\begin{aligned}
\bm\Theta_{q,x}^{(t)}(\boldsymbol x,\omega_q,\iota_q)
\!\triangleq\!  
\begin{bmatrix}
\omega_q\mathbf{I}_M-\bm\Xi & -\boldsymbol c_q^{(t)}(\boldsymbol x)\\
-\boldsymbol c_q^{(t)}(\boldsymbol x)^H & \eta\sigma_q^2-\iota_q-\omega_q e_q^2
\end{bmatrix}\!\succeq\!\boldsymbol 0,\\
\forall q\in\mathcal Q,
\end{aligned}
\end{equation}
where $\iota_q$ is an auxiliary variable satisfying 
\begin{equation}\label{eq:48}
\iota_q \ge \upsilon_q^{(t)}(\boldsymbol x).
\end{equation}

Based on the above derivations, a convex approximation of problem~\eqref{p3_raw} is formulated as
\begin{subequations}\label{p3_sca}
\begin{align}
\mathmakebox[2em][r]{\underset{\mathclap{\boldsymbol x,\tau,\{\omega_q,\iota_q\}_{q\in\mathcal Q}}}{\max}}\;
& \quad\tau
\\
\mathmakebox[2em][r]{\text{s.t.}}\;
&\quad (\ref{Const_4}),(\ref{Const_5}),\eqref{eq:sinr_sca_constraint_position},\eqref{eq:position_covert_sca_lmi_relaxed},\eqref{eq:48},
\end{align}
\end{subequations}
which is convex and can be optimally solved using the CVX toolbox.
\begin{proposition}\phantomsection\label{prop:bcd_wc_monotonicity}
The value of the objective function in Problem~\eqref{p3_sca} is non-decreasing with iterations, and Algorithm~\ref{alg:robust_bcd} converges.
\end{proposition}
\noindent\textit{Proof.} Please refer to Appendix~\ref{AppBCDWC}.
\section{Outage-Constrained Robust Beamforming under Statistical CSI Errors}\label{sec:stat_csi_design}
In practice, the worst-case optimization scheme may be overly conservative, since it requires the covertness constraint to hold for all CSI error realizations in the uncertainty set regardless of their statistical distribution. To address this issue, in this section, we focus on the statistical warden CSI error model and investigate the corresponding outage-constrained robust beamforming optimization. In the following, we first formulate the design problem and derive a deterministic approximation to the outage-constrained covertness constraint using the Bernstein-type inequality.
Then, we adopt the BCD framework to alternately optimize the beamforming, AN covariance matrix, and MA positions.
\subsection{Problem Formulation}
As in the worst-case design, we lift the transmit beamforming vectors into matrix variables. Let $\tau$ denote the SINR threshold. The optimization problem to maximize the minimum weighted SINR among all legitimate users is formulated as
\begin{subequations}\label{p2_stat}
\begin{align}
\mathmakebox[2em][r]{\underset{\mathclap{\tau,\boldsymbol W_k,\boldsymbol{R},\boldsymbol x}}{\max}}\; & \quad\tau
\\
\mathmakebox[2em][r]{\text{s.t.}}\;
&\quad
\frac{\boldsymbol h_k^H(\boldsymbol{x}) \boldsymbol{W}_k \boldsymbol h_k(\boldsymbol{x})}
{\boldsymbol h_k^H(\boldsymbol{x}) (\boldsymbol{R}\!-\!\boldsymbol W_k)\boldsymbol h_k(\boldsymbol{x})+\sigma_k^2}
\ge \nu_k\tau,\forall k\in\mathcal K,
\label{Const_stat_sinr}\\
&\quad\mathrm{Pr}\!\left\{
{\boldsymbol h}_q(\boldsymbol x)^H
\bm\Xi
{\boldsymbol h}_q(\boldsymbol x)
\le \eta\sigma_q^2
\right\}
\ge 1-\varrho_q,\notag\\
&\quad\Delta\boldsymbol h_q \sim\mathcal{CN}(\boldsymbol 0,\boldsymbol E_q),
\forall q\in\mathcal Q,
\label{Const_stat_covert}\\
&\quad\mathrm{Tr}(\boldsymbol{R})\le P,
\label{Const_stat_power}\\
&\quad\boldsymbol{R}-\sum_{k\in\mathcal K}\boldsymbol W_k\succeq\boldsymbol 0,
\boldsymbol W_k\succeq\boldsymbol 0,
\forall k\in\mathcal K,
\label{Const_stat_RA_split}\\
&\quad\mathrm{rank}(\boldsymbol W_k)=1,
\quad \forall k\in\mathcal K,
\label{Const_stat_rank}\\
&\quad\eqref{Const_4},\eqref{Const_5}.
\end{align}
\end{subequations}
where $\varrho_q\in(0,1)$ denotes the maximum tolerable outage probability for the covertness requirement at warden $q$. Constraint \eqref{Const_stat_covert} represents the outage-constrained covertness requirement, ensuring that \eqref{eq:covert_compact} holds with probability at least $1-\varrho_q$. The probabilistic quadratic constraint~\eqref{Const_stat_covert} renders problem~\eqref{p2_stat} intractable.
\subsection{Optimizing Beamforming and AN Covariance Matrix}
To obtain a tractable formulation of \eqref{Const_stat_covert}, we use the following Bernstein-type inequality.
\begin{lemma}[Bernstein-Type Inequality]\phantomsection\label{lem:BTI_statCSI}
\it For any complex Hermitian matrix $\mathbf{A}\in\mathbb{C}^{M\times M}$, vector $\mathbf{b}\in\mathbb{C}^{M\times1}$, scalar $c\in\mathbb{R}$, random vector $\mathbf{x}\sim\mathcal{CN}(\bm0_{M\times1},\mathbf{I}_M)$, and $\rho\in(0,1)$, if there exist $x$ and $y$ such that
\begin{equation}\label{eq:BTI_generic}
\begin{aligned}
&{\rm Tr}(\mathbf{A}) - \sqrt {2\ln \left(\frac{1}{\rho }\right)}x  + \ln(\rho )y + c \ge 0,\\&
\sqrt {\left\|\mathbf{A} \right\|_F^2 + 2{{\left\| \mathbf{b} \right\|}^2}}  \le x, \\&
y\mathbf{I}_M + \mathbf{A} \succeq \boldsymbol 0,y \ge 0,
\end{aligned}
\end{equation}
then
\begin{equation}
{\rm Pr}({\mathbf{x}^H}{\mathbf{A}}\mathbf{x} + 2{\mathop{\rm Re}\nolimits} \{ \mathbf{x}^H\mathbf{b}\}  + {c} \ge 0)\ge 1-\rho.
\end{equation}
\end{lemma}
Let $\boldsymbol d_q\sim\mathcal{CN}(\boldsymbol 0_{M \times 1},\mathbf{I}_M)$ and
represent the CSI error as
$\Delta\boldsymbol h_q=\boldsymbol E_q^{1/2}\boldsymbol d_q$,
which is valid since $\boldsymbol E_q\succeq\boldsymbol{0}_{M\times M}$ admits a
Hermitian positive-semidefinite square root $\boldsymbol E_q^{1/2}$.
Then, \eqref{Const_stat_covert} can be expressed as
\eqref{eq:transform_outage} at the bottom of this page.
\begin{figure*}[!b]
\hrulefill
\begin{equation}\label{eq:transform_outage}
\begin{aligned}
&\mathrm{Pr}\Bigl\{
- \left(\bar{\boldsymbol h}_q(\boldsymbol x)+\boldsymbol E_q^{1/2}\boldsymbol d_q\right)^H
\bm\Xi
\left(\bar{\boldsymbol h}_q(\boldsymbol x)+\boldsymbol E_q^{1/2}\boldsymbol d_q\right)
+\eta\sigma_q^2 \ge0
\Bigr\}\\
&=\mathrm{Pr}\Bigl\{-\boldsymbol d_q^H\left(\boldsymbol E_q^{1/2}\right)^H\bm\Xi\boldsymbol E_q^{1/2}\boldsymbol d_q
-2\Re\left\{\bar{\boldsymbol h}_q^H(\boldsymbol x)\bm\Xi\boldsymbol E_q^{1/2}\boldsymbol d_q\right\}
+\eta\sigma_q^2-
\bar{\boldsymbol h}_q^H(\boldsymbol x)\bm\Xi\bar{\boldsymbol h}_q(\boldsymbol x)\ge0\Bigr\}
\end{aligned}
\end{equation}
\end{figure*}
Therefore, the outage-constrained covertness requirement \eqref{Const_stat_covert} is given by
\begin{equation}\label{eq:stat_quad_form}
\mathrm{Pr}\left\{
\boldsymbol d_q^H\boldsymbol U_q\boldsymbol d_q
\!+\!2\Re\{\boldsymbol u_q^H(\boldsymbol x)\boldsymbol d_q\}
\!+\!u_q(\boldsymbol x)
\!\ge\!0
\right\}
\!\ge\!1\!-\!\varrho_q,
\end{equation}
where
\begin{subequations}
\begin{align}
&\boldsymbol U_q
\triangleq
-\left(\boldsymbol E_q^{1/2}\right)^H\bm\Xi\boldsymbol E_q^{1/2},\label{eq:stat_Uq_def}\\
&\boldsymbol u_q(\boldsymbol x)
\triangleq
-\left(\boldsymbol E_q^{1/2}\right)^H\bm\Xi\bar{\boldsymbol h}_q(\boldsymbol x),\label{eq:stat_uq_def}\\
&u_q(\boldsymbol x)
\triangleq
\eta\sigma_q^2-
\bar{\boldsymbol h}_q^H(\boldsymbol x)\bm\Xi\bar{\boldsymbol h}_q(\boldsymbol x) = 
\eta\sigma_q^2-\varkappa_q(\boldsymbol x).\label{eq:stat_scalar_def}
\end{align}
\end{subequations}

According to Lemma~\ref{lem:BTI_statCSI}, sufficient deterministic conditions for the outage-constrained covertness requirement in \eqref{eq:stat_quad_form} are given by
\begin{subequations}
\begin{align}
&\mathrm{Tr}(\boldsymbol U_q)
\!-\!\sqrt{2\ln(1/\varrho_q)}\,\ell_q
\!-\!\ln(1/\varrho_q)\,\psi_q
\!+\!u_q(\boldsymbol x)
\ge0,\label{eq:stat_covert_BTI_a}\\
&\sqrt{\|\boldsymbol U_q\|_F^2+2\|\boldsymbol u_q(\boldsymbol x)\|_2^2}
\le \ell_q,\label{eq:stat_covert_BTI_b}\\
&\psi_q\mathbf{I}_M+\boldsymbol U_q\succeq\boldsymbol 0,
\psi_q\ge0,\label{eq:stat_covert_BTI_c}
\end{align}
\end{subequations}
where $\boldsymbol{\ell}=[\ell_1,\ldots,\ell_Q]^{\mathrm T}$ and $\boldsymbol{\psi}=[\psi_1,\ldots,\psi_Q]^{\mathrm T}$ are vectors of auxiliary variables. Due to the affine dependence of $\bm\Xi$ on $\boldsymbol W_k$ and $\boldsymbol{R}$, constraints \eqref{eq:stat_covert_BTI_a}--\eqref{eq:stat_covert_BTI_c} are convex in the beamforming covariance matrices when the antenna positions are fixed.

Therefore, for a fixed SINR threshold $\tau$ and a fixed APV $\boldsymbol x$, the beamforming feasibility problem under statistical CSI errors is formulated as
\begin{subequations}\label{p2_stat_feas}
\begin{align}
\mathmakebox[2em][r]{\text{find}}\;
&\quad \boldsymbol W_k,\boldsymbol{R},\boldsymbol{\ell},\boldsymbol{\psi}
\\
\mathmakebox[2em][r]{\text{s.t.}}\;
&\quad\eqref{Const_stat_sinr},\eqref{eq:stat_covert_BTI_a},\eqref{eq:stat_covert_BTI_b},\eqref{eq:stat_covert_BTI_c},\eqref{Const_stat_power},\eqref{Const_stat_RA_split},\eqref{Const_stat_rank}.
\end{align}
\end{subequations}
Similarly, by dropping the rank-one constraint in \eqref{Const_stat_rank}, Problem~\eqref{p2_stat_feas} is transformed into an SDP with second-order cone constraints and can be solved by standard convex solvers. 

Note that, for a feasible solution with $\tau>0$, the rank-one recovery in \eqref{SDR_optimal} preserves the user SINRs, total covariance matrix, and original outage-constrained covertness constraint.

\subsection{Optimizing MA Positions}\label{subsec:MA_position_update_statCSI}
In each outer iteration, we retain the relaxed covariance matrices $\{\overline{\boldsymbol W}_k\}_{k\in\mathcal K}$ and $\overline{\boldsymbol R}$, together with their feasible BTI auxiliary variables, after rank-one recovery. The recovered beamforming vectors and AN covariance matrix are held fixed for evaluating the user SINRs. At each inner SCA iteration $t$, the auxiliary variables $\alpha_k^{(t)}$ are updated via \eqref{eq:alpha_update_position} at the current position $\boldsymbol x^{(t)}$, using $\boldsymbol w_k=\boldsymbol w_k^\star$ and $\boldsymbol\Xi_k=\overline{\boldsymbol R}-\boldsymbol W_k^\star$. The SINR constraint is then handled by \eqref{eq:sinr_sca_constraint_position}.

For the covertness constraints in this subsection, we use the fixed matrix $\overline{\bm\Xi}\triangleq(1+\eta)\sum_{k\in\mathcal K_c}\overline{\boldsymbol W}_k-\eta\overline{\boldsymbol R}$. Specifically, $\bm\Xi$ is replaced by $\overline{\bm\Xi}$ in \eqref{eq:stat_Uq_def}--\eqref{eq:stat_scalar_def}, in $\varkappa_q(\boldsymbol x)$, and in its upper bound \eqref{eq:varkappa_q_upper_bound} and Hessian bound in Appendix~\ref{AppSINR_a}. All covertness approximations below use this same matrix throughout the inner SCA loop. Note that $\boldsymbol U_q$ is independent of $\boldsymbol x$, whereas $\boldsymbol u_q(\boldsymbol x)$ and $u_q(\boldsymbol x)$ depend on $\boldsymbol x$ through $\bar{\boldsymbol h}_q(\boldsymbol x)$.
For constraint \eqref{eq:stat_covert_BTI_b}, define
\begin{equation}\label{eq:varsigma_def}
\varsigma_q(\boldsymbol x)
\triangleq
\|\boldsymbol u_q(\boldsymbol x)\|_2^2
=
\bar{\boldsymbol h}_q^H(\boldsymbol x)\,
\boldsymbol A_q\,
\bar{\boldsymbol h}_q(\boldsymbol x),
\end{equation}
where $\boldsymbol A_q\triangleq\overline{\bm\Xi}\boldsymbol E_q\overline{\bm\Xi}
=\left(\boldsymbol E_q^{1/2}\overline{\bm\Xi}\right)^H\!\left(\boldsymbol E_q^{1/2}\overline{\bm\Xi}\right)
\succeq\boldsymbol 0$ is a constant Hermitian matrix with the beamforming
covariance matrices fixed. Since $\varsigma_q(\boldsymbol x)$ has the same
quadratic form as $\varkappa_q(\boldsymbol x)$, its Hessian bound follows
from Appendix~\ref{AppSINR_a} by replacing $\bm\Xi$ with
$\boldsymbol A_q$, i.e.,
$\nabla_{\boldsymbol{x}}^2 \varsigma_q(\boldsymbol{x})
\preceq\varpi_{\varsigma,q}\mathbf{I}_M$ with
\begin{equation}
\varpi_{\varsigma,q}
=
\frac{16\pi^2 (M-1)}{\lambda^2}|\beta_q|^2\max_{m,n\in\mathcal{M}}\left|[\boldsymbol A_q]_{m,n}\right|.
\end{equation}
Based on the above Hessian bound, a global quadratic upper bound on $\varsigma_q(\boldsymbol x)$ is given by
\begin{equation}\label{eq:varsigma_q_upper_bound}
\begin{aligned}
\varsigma_q(\boldsymbol x)\le\,
\varsigma_q(\boldsymbol{x}^{(t)})
& + \nabla_{\boldsymbol{x}}\varsigma_q(\boldsymbol{x}^{(t)})^T
(\boldsymbol{x}-\boldsymbol{x}^{(t)})
\\&
+ \frac{\varpi_{\varsigma,q}}{2}
\|\boldsymbol{x}-\boldsymbol{x}^{(t)}\|_2^2 \triangleq \bar{\varsigma}_q^{(t)}(\boldsymbol{x}),
\end{aligned}
\end{equation}
where equality holds at $\boldsymbol x=\boldsymbol x^{(t)}$.
For $\varpi_{\varsigma,q}>0$, the quadratic upper bound in \eqref{eq:varsigma_q_upper_bound} can be equivalently rewritten as
$\bar{\varsigma}_q^{(t)}(\boldsymbol{x})
=\frac{\varpi_{\varsigma,q}}{2}\|\boldsymbol x-\boldsymbol v_q^{(t)}\|_2^2
+d_q^{(t)}$, where
$\boldsymbol v_q^{(t)}
\triangleq
\boldsymbol x^{(t)}-\frac{1}{\varpi_{\varsigma,q}}
\nabla_{\boldsymbol{x}}\varsigma_q(\boldsymbol{x}^{(t)})$ and
$d_q^{(t)}
\triangleq
\varsigma_q(\boldsymbol{x}^{(t)})
-\frac{\|\nabla_{\boldsymbol{x}}\varsigma_q(\boldsymbol{x}^{(t)})\|_2^2}
{2\varpi_{\varsigma,q}}$.
Moreover, since \eqref{eq:varsigma_q_upper_bound} holds for all $\boldsymbol x$
and $\varsigma_q(\boldsymbol x)\ge0$, we have
$d_q^{(t)}=\min_{\boldsymbol x}\bar{\varsigma}_q^{(t)}(\boldsymbol{x})
\ge\varsigma_q(\boldsymbol v_q^{(t)})\ge0$.
Therefore, a sufficient second-order cone condition for
constraint \eqref{eq:stat_covert_BTI_b} is given by
\begin{equation}\label{eq:stat_position_BTI_b_soc}
\left\|
\left[
\|\boldsymbol U_q\|_F,\,
\sqrt{\varpi_{\varsigma,q}}\,(\boldsymbol x-\boldsymbol v_q^{(t)})^T,\,
\sqrt{2d_q^{(t)}}
\right]^T
\right\|_2
\le \ell_q,
\forall q\in\mathcal Q.
\end{equation}
Note that if $\varpi_{\varsigma,q}=0$, $\varsigma_q(\boldsymbol x)$ is constant, and \eqref{eq:stat_position_BTI_b_soc} is replaced by $\sqrt{\|\boldsymbol U_q\|_F^2+2\varsigma_q(\boldsymbol x^{(t)})}\le\ell_q$.
For constraint \eqref{eq:stat_covert_BTI_a}, by applying the upper bound in \eqref{eq:varkappa_q_upper_bound}, we have
\begin{equation}\label{eq:stat_position_BTI_a_sca}
\begin{aligned}
&\mathrm{Tr}(\boldsymbol U_q)
-\sqrt{2\ln(1/\varrho_q)}\,\ell_q
-\ln(1/\varrho_q)\,\psi_q\\
&\quad+\eta\sigma_q^2-\bar{\varkappa}_q^{(t)}(\boldsymbol x)
\ge0,
\quad\forall q\in\mathcal Q.
\end{aligned}
\end{equation}
Thus, the resulting subproblem for MA position optimization can be formulated as
\begin{subequations}\label{p3_stat_sca}
\begin{align}
\mathmakebox[2em][r]{\underset{\mathclap{\boldsymbol x,\tau,\boldsymbol{\ell},\boldsymbol{\psi}}}{\max}}\; &\quad
\tau
\\
\mathmakebox[2em][r]{\text{s.t.}}\;
&\quad\eqref{eq:sinr_sca_constraint_position},
\eqref{eq:stat_position_BTI_a_sca},
\eqref{eq:stat_position_BTI_b_soc},
\eqref{eq:stat_covert_BTI_c},\eqref{Const_4},\eqref{Const_5},
\end{align}
\end{subequations}
which is convex and can be optimally solved using the CVX toolbox.


\subsection{Algorithm Framework and Complexity Analysis}\label{subsec:algorithm_complexity}
The overall BCD procedure is summarized in Algorithm~\ref{alg:robust_bcd}.
The bisection search is detailed in Algorithm~\ref{alg:robust_bisection}. 
We next analyze the complexity of the convex subproblems using the LMI and SOC counting method in~\cite{WangOutageConic2014}. The beamforming and AN covariance matrix subproblem has complexity $\mathcal C_{{\rm B},{\rm WC}}=\mathcal O\bigl({(K+Q)^{1.5}}\allowbreak M^{2.5}\allowbreak{(KM^2+Q)^2}\bigr)$ for the worst-case optimization scheme and $\mathcal C_{{\rm B},{\rm ST}}=\mathcal O\bigl({(K+Q)^{1.5}}\allowbreak M^{2.5}\allowbreak{(KM^2+2Q)^2}\bigr)$ for the statistical optimization scheme. The MA position subproblem has complexity $\mathcal C_{{\rm X},{\rm WC}}=\mathcal O\bigl(M^2(M+2Q)\allowbreak(QM+K)^{0.5}\allowbreak(QM+2Q^2+K)\bigr)$ for the worst-case optimization scheme and $\mathcal C_{{\rm X},{\rm ST}}=\mathcal C_{{\rm X},{\rm WC}}$ for the statistical optimization scheme.

To assess the computational cost in practice, we conduct a runtime test using the default parameters in Section~V. The test is performed on a computer equipped with an Intel Core i7-12700 processor. The average runtimes per outer BCD iteration are $3.42$ s and $4.17$ s for the worst-case and statistical optimization schemes, respectively.
\begin{algorithm}[!t]
\small
\SetAlgoNoLine
\SetNlSty{textnormal}{}{:\,}
\SetAlgoNlRelativeSize{0}
\SetAlgoCaptionSeparator{}
\SetAlCapHSkip{0pt}
\SetAlgoHangIndent{0pt}
\SetArgSty{textnormal}
\SetKwInput{KwInit}{Initialize}
\caption{BCD Algorithm for Solving Problems \eqref{p1} and \eqref{p2_stat}}
\label{alg:robust_bcd}
\KwInit{$\boldsymbol x^{(r)}$, $\{\boldsymbol w_k^{(r)}\}$, $\boldsymbol R_{\rm A}^{(r)}$, and set $r=0$.}

\SetEndCharOfAlgoLine{.}
\Repeat{exit conditions are met}{
    \SetEndCharOfAlgoLine{;}
    Update $\tau_{\rm B}^{(r+1)}$ by the bisection search in Algorithm~\ref{alg:robust_bisection}\;
    Update $\overline{\boldsymbol W}_k^{(r+1)}$ and $\overline{\boldsymbol R}^{(r+1)}$ via \eqref{p2} or \eqref{p2_stat_feas} at $\tau_{\rm B}^{(r+1)}$\; 
    Construct $\boldsymbol w_k^{(r+1)}$ and $\boldsymbol R_{\rm A}^{(r+1)}$ via \eqref{SDR_optimal}\;
    Update $\alpha_k^{(r+1)}$ via \eqref{eq:alpha_update_position}\;
    Update $\boldsymbol x^{(r+1)}$ and $\tau^{(r+1)}$ via \eqref{p3_sca} or \eqref{p3_stat_sca}\;
    Update $r=r+1$\;
}
\KwRet{$\boldsymbol x^\star$, $\boldsymbol w_k^\star$, and $\boldsymbol R_{\rm A}^\star$}\;
\end{algorithm}

\begin{algorithm}[!t]
\small
\SetAlgoNoLine
\SetNlSty{textnormal}{}{:\,}
\SetAlgoNlRelativeSize{0}
\SetAlgoCaptionSeparator{}
\SetAlCapHSkip{0pt}
\SetAlgoHangIndent{0pt}
\SetArgSty{textnormal}
\SetKwInput{KwInit}{Initialize}
\caption{Bisection Search Algorithm}
\label{alg:robust_bisection}
\KwInit{$\boldsymbol x^{(r)}$, $\epsilon_{\rm b}$, $\tau_{\rm L}=\tau^{(r)}$, and $\tau_{\rm U}=\min_{k\in\mathcal K}\frac{P\|\boldsymbol h_k(\boldsymbol x^{(r)})\|_2^2}{\nu_k\sigma_k^2}$.}
\SetEndCharOfAlgoLine{.}
\Repeat{$\tau_{\rm U}-\tau_{\rm L}\le\epsilon_{\rm b}$}{
    \SetEndCharOfAlgoLine{;}
    Update $\tau=(\tau_{\rm L}+\tau_{\rm U})/2$\;
    Solve problem~\eqref{p2} or \eqref{p2_stat_feas} with fixed $\boldsymbol x^{(r)}$ and $\tau$\;
    \eIf{the problem is feasible}{
        Update $\tau_{\rm L}=\tau$\;
    }{
        Update $\tau_{\rm U}=\tau$\;
    }
}
\KwRet{$\tau_{\rm B}^{(r+1)}=\tau_{\rm L}$}\;
\end{algorithm}


\section{Simulation Results}
This section presents numerical results to assess the effectiveness of the proposed algorithms for the MA-assisted covert communication system. Unless otherwise specified, $M=6$, $K_o=K_c=Q=2$, $P=0$ dBW, and the noise power at all users and wardens is $-80$ dBW. Alice is located at the origin of a $60$ m $\times$ $60$ m field, and all users and wardens have a single fixed antenna. Overt users 1 and 2 are randomly placed in disks of radius $r_o=15$ m centered $60$ m from Alice at $70^\circ$ and $110^\circ$, respectively. Each disk contains one covert user and one warden. The linear transmit region has length $A=12\lambda$, with $\lambda=0.1$ m and minimum MA spacing $x_{th}=\lambda/2$. Each user channel contains one LoS path and $L_k-1$ non-line-of-sight (NLoS) paths, with $L_k=3$. The independent path gains satisfy $\beta_k^1\sim\mathcal{CN}\bigl(0,{\kappa c_k^2}/(\kappa+1)\bigr)$ for the LoS path and $\beta_k^\ell\sim\mathcal{CN}\bigl(0,{c_k^2}/{((\kappa+1)(L_k-1))}\bigr)$ for the NLoS paths, where $\ell=2,\ldots,L_k$ and $\kappa=2$. The large-scale path loss is $c_k^2=C_0d_k^{-\alpha_{\mathrm{PL}}}$, where $d_k$ is the link distance in meters, $C_0=-30$ dB, and $\alpha_{\mathrm{PL}}=2.6$. Similarly, the warden channel gain is $\beta_q=c_qe^{\jmath\phi_q}$, where $\phi_q\sim\mathcal{U}[0,2\pi)$, $c_q^2=C_0d_q^{-\alpha_{\mathrm{PL}}}$, and $d_q$ is the link distance in meters. The LoS azimuth angles are determined by the user and warden locations, while the NLoS azimuth angles are independently and uniformly distributed over $[0,\pi]$. The transmission blocklength is $N=32$, the covertness level is $\epsilon=0.1$, and the covert user priority parameter is $\nu=0.5$.
For a fair comparison between the worst-case and statistical optimization schemes, the CSI uncertainty coefficient is set to $\kappa_{\mathrm{CSI}}=0.01$ and the maximum covertness outage probability is $\varrho_q=0.05$ for every warden. Following the relative error model in~\cite{ZhangRobustCovertISAC}, the statistical error covariance matrix is set to $\boldsymbol E_q=\kappa_{\mathrm{CSI}}\|\bar{\boldsymbol h}_q(\boldsymbol x)\|_2^2\boldsymbol I_M/M$. Accordingly, the bounded-error radius is chosen as $e_q^2=\frac{\kappa_{\mathrm{CSI}}\|\bar{\boldsymbol h}_q(\boldsymbol x)\|_2^2}{2M}F_{\chi_{2M}^2}^{-1}(1-\varrho_q)$, where $F_{\chi_{2M}^2}^{-1}$ denotes the inverse cumulative distribution function of a chi-square distribution with $2M$ DoFs. 
\begin{figure}[!t]
  \centering
  \includegraphics[width=\columnwidth]{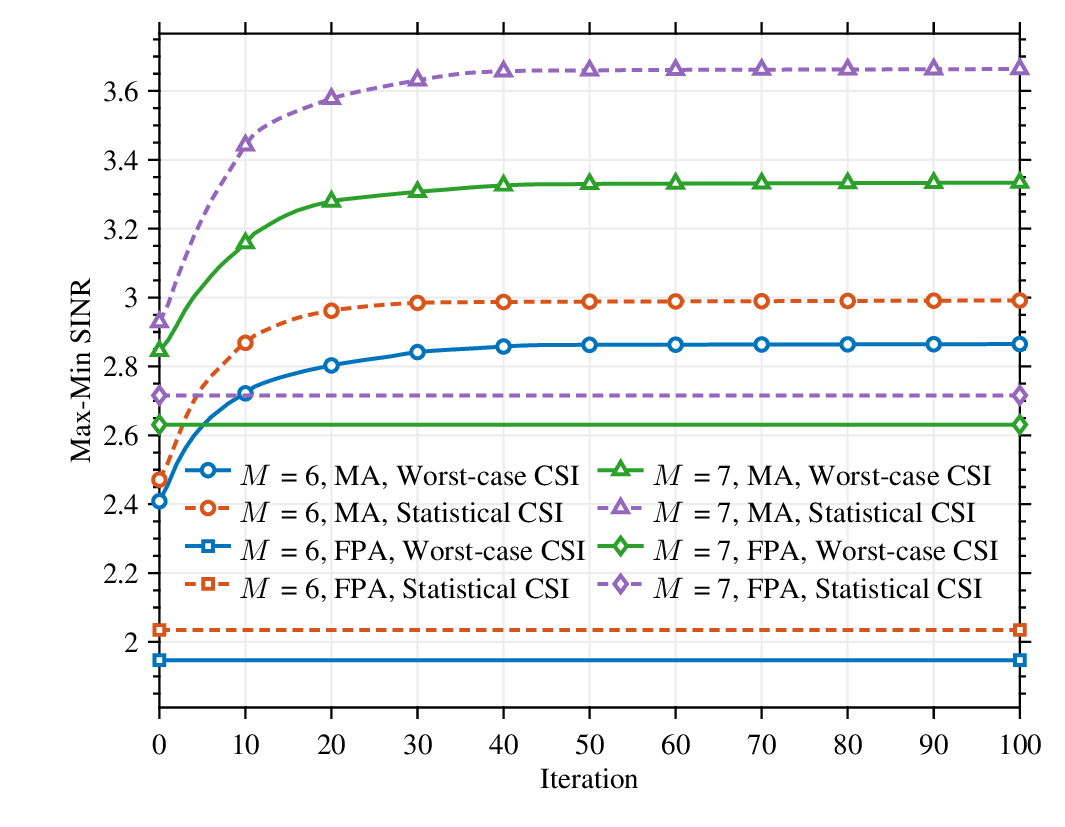}
  \caption{Convergence behavior of the proposed algorithms and FPA schemes.}
  \label{fig:bcd_iteration_m}
\end{figure}

Fig.~\ref{fig:bcd_iteration_m} depicts the convergence behavior of the proposed MA algorithms and the  corresponding FPA schemes under different CSI error models. It can be observed that the weighted max-min SINR achieved by the proposed MA algorithms increases monotonically and converges repidly after 30 iterations. Note that the number of antennas $M$ has a minor impact on the convergence rate.
In terms of weighted max-min SINR, the proposed MA algorithms outperform the corresponding FPA schemes under both CSI error models. This superior performance is attributed to the active movement of antenna elements, which enhances the desired user channels and provides additional spatial DoFs to suppress covert signal leakage toward wardens.

\begin{figure}[!t]
  \centering
  \includegraphics[width=\columnwidth]{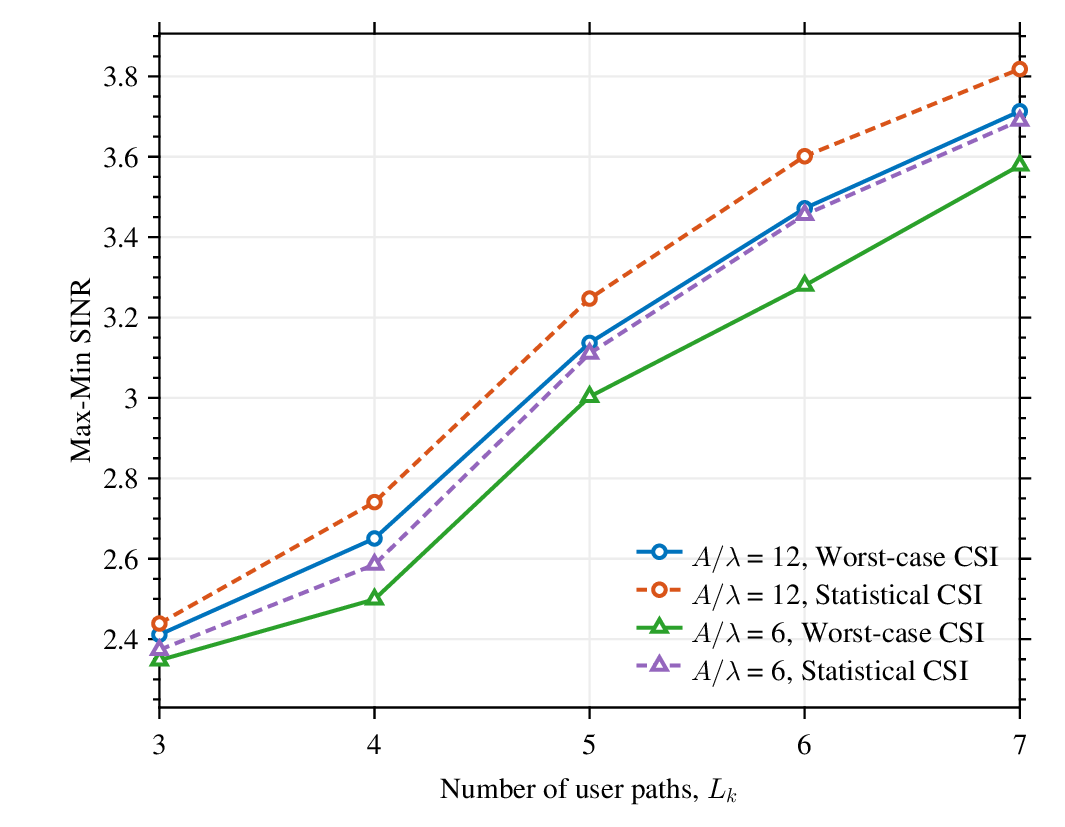}
  \caption{Weighted max-min SINR versus the number of user paths.}
  \label{fig:path_number}
\end{figure}

Fig.~\ref{fig:path_number} illustrates the max-min SINR versus the number of user paths for different movable region sizes under the bounded and statistical CSI error models. 
The azimuth angles from Alice to the centers of the two covert user regions are $65^\circ$ and $95^\circ$, respectively, and the azimuth angles from Alice to the two wardens are $40^\circ$ and $120^\circ$, respectively. It can be observed that the max-min SINR increases with increasing $L_k$ under both CSI error models. This is because antenna movement can exploit the spatial diversity provided by additional propagation paths to enhance the desired channel conditions. In addition, the schemes with $A/\lambda=12$ outperform their counterparts with $A/\lambda=6$, 
as a larger movable region offers more flexibility in adjusting the antenna positions. 

\begin{figure}[!t]
  \centering
  \includegraphics[width=\columnwidth]{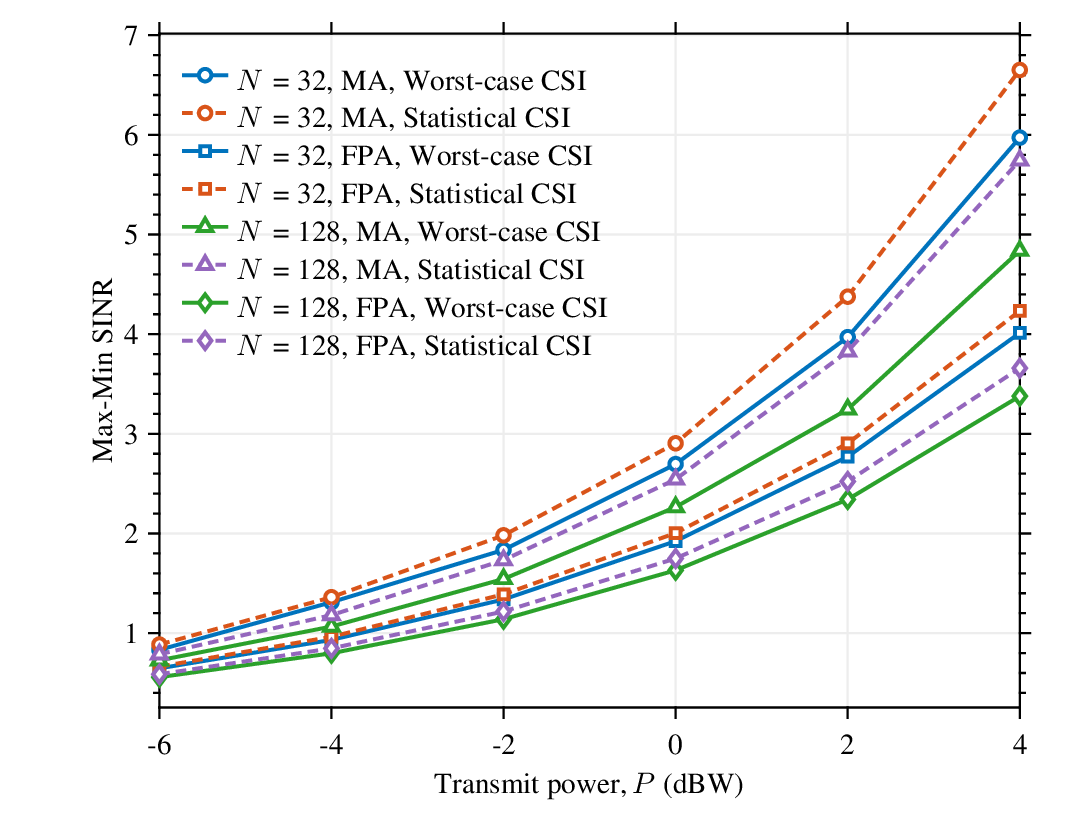}
  \caption{Weighted max-min SINR versus the maximum transmit power.}
  \label{fig:transmit_power}
\end{figure}

Fig.~\ref{fig:transmit_power} illustrates the weighted max-min SINR versus the maximum transmit power $P$ for different transmission blocklengths.
It can be observed that the weighted max-min SINRs of all schemes increase as $P$ increases. 
This is because a larger transmit-power budget provides more power for the transmit signal while satisfying the covertness constraints. Moreover, the proposed MA schemes outperform the corresponding FPA schemes by properly optimizing the antenna positions to enhance the desired channels and mitigate interference. It can also be observed that the weighted max-min SINR achieved with $N=128$ is lower than that with $N=32$.
This is because a larger $N$ provides each warden with more observations, improving the accuracy of received power estimation and the reliability of covert transmission detection.
For a fixed received power ratio, the KL divergence in \eqref{eq:KL_div} increases linearly with $N$.
For a fixed $\epsilon$, the threshold $\eta$ therefore decreases, imposing more stringent constraints on covert signal leakage.
These constraints restrict the feasible region and limit the achievable weighted max-min SINR.

\begin{figure}[!t]
  \centering
  \includegraphics[width=\columnwidth]{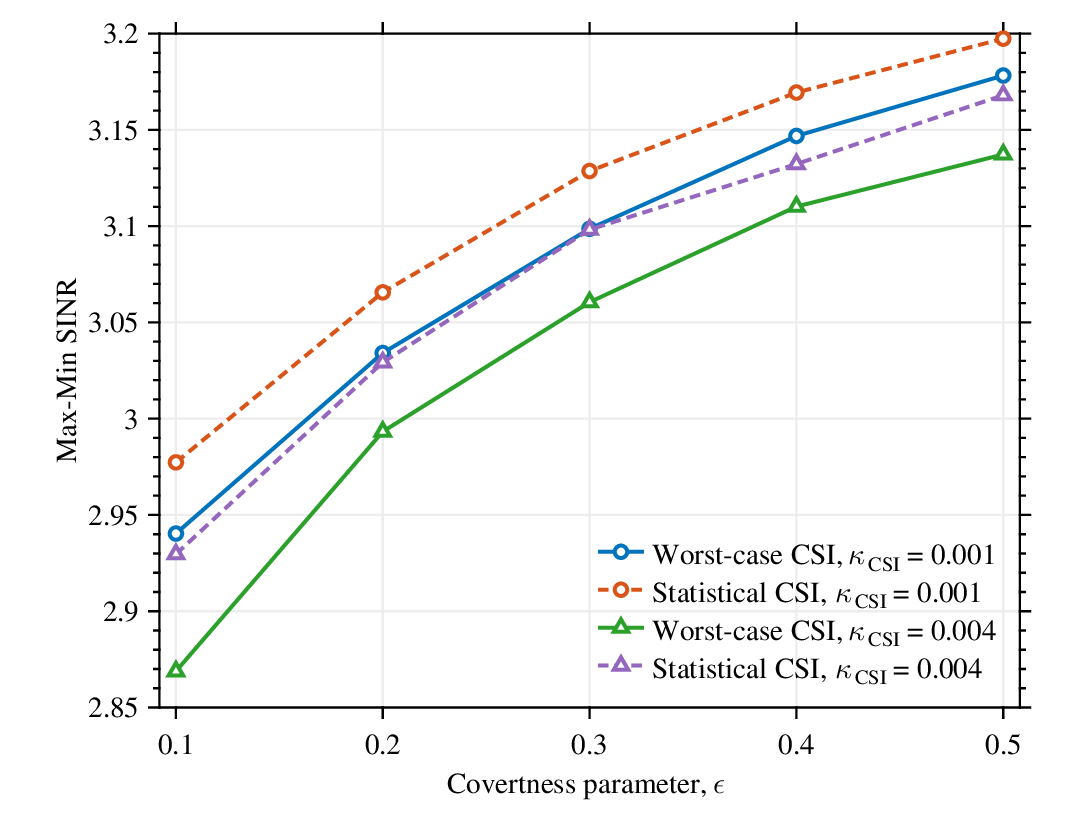}
  \caption{Weighted max-min SINR versus the covertness level for different CSI uncertainty levels.}
  \label{fig:epsilon_kappa_csi}
\end{figure}

In Fig.~\ref{fig:epsilon_kappa_csi}, we present the weighted max-min SINR against the covertness level $\epsilon$ under different relative CSI uncertainty levels. The azimuth angles from Alice to the centers of the two user regions are $60^\circ$ and $95^\circ$, respectively, and the azimuth angles from Alice to the two wardens are $80^\circ$ and $115^\circ$, respectively. It can be observed that the weighted max-min SINR increases as $\epsilon$ increases under both CSI error models. As $\epsilon$ increases, the covertness constraint is relaxed. Consequently, the feasible region of optimization becomes larger and a higher weighted max-min SINR can be achieved. It is also observed that the weighted max-min SINR becomes lower as $\kappa_{\mathrm{CSI}}$ increases from $0.001$ to $0.004$. This is because, with a fixed covertness outage probability, a larger $\kappa_{\mathrm{CSI}}$ increases both the error variance and the bounded-error radius. 

\begin{figure}[!t]
  \centering
  \includegraphics[width=\columnwidth]{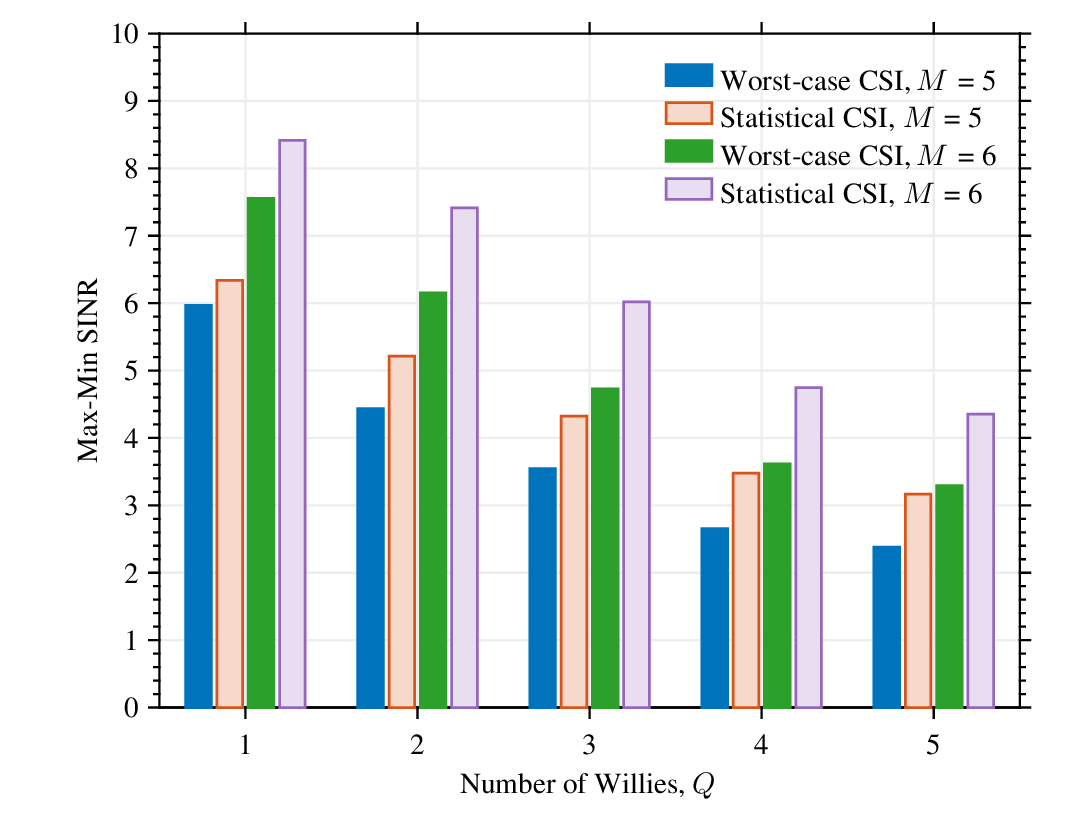}
  \caption{Weighted max-min SINR versus the number of wardens.}
  \label{fig:willie_number}
\end{figure}

Fig.~\ref{fig:willie_number} illustrates the effect of the number of wardens $Q$ on the weighted max-min SINR under the two CSI error models. To more clearly illustrate the impact of increasing $Q$, we consider a scenario with one overt user and one covert user. The azimuth angle from Alice to the center of the user region is $90^\circ$. It can be observed that the weighted max-min SINR decreases as $Q$ increases. This is because each additional warden attempts to detect the covert transmission, making the overall covertness requirements stricter. It can also be observed that, for the same number of antennas, the statistical optimization scheme achieves higher weighted max-min SINR than the worst-case scheme.
For instance, at $Q=3$, the statistical scheme outperforms the worst-case scheme by approximately $0.86$ dB and $1.05$ dB for $M=5$ and $M=6$, respectively.
This is because the worst-case scheme requires the covertness constraints to hold for all channel errors within the uncertainty set.
These more stringent constraints reduce the flexibility in power allocation and antenna positioning, leading to a lower weighted max-min SINR.

\begin{figure}[!t]
  \centering
  \includegraphics[width=\columnwidth]{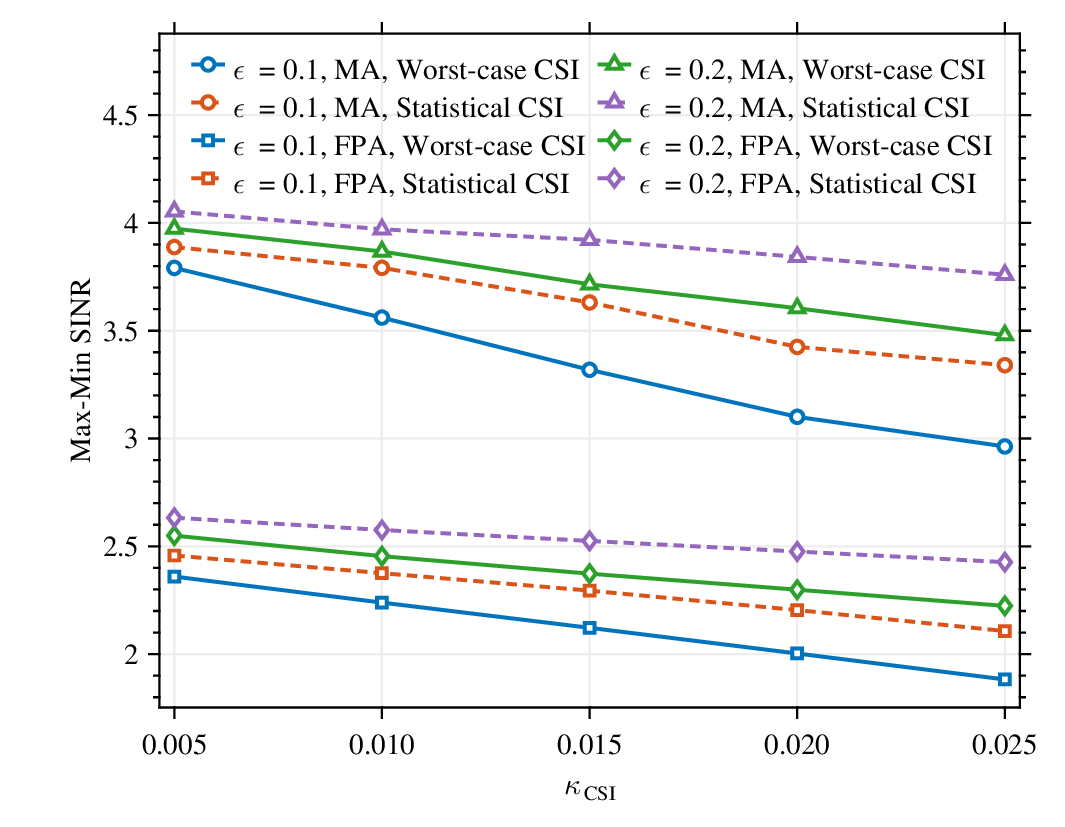}
  \caption{Weighted max-min SINR versus relative CSI uncertainty for equivalent covertness levels.}
  \label{fig:kappa_csi_equiv_epsilon}
\end{figure}

In Fig.~\ref{fig:kappa_csi_equiv_epsilon}, we present the weighted max-min SINR versus the relative CSI uncertainty $\kappa_{\mathrm{CSI}}$ for $\epsilon=0.1$ and $0.2$, where $\kappa_{\mathrm{CSI}}$ ranges from $0.005$ to $0.025$. It can be observed that the weighted max-min SINR of all schemes decreases with increasing $\kappa_{\mathrm{CSI}}$. This is because increased uncertainty in the warden channels makes it more difficult to satisfy the covertness constraints. For the same $\epsilon$, the proposed MA schemes outperform the corresponding FPA schemes by properly optimizing the antenna positions to enhance the desired user channels and suppress information leakage. In addition, the statistical optimization scheme achieves higher weighted max-min SINR than the worst-case optimization scheme for both MA and FPA setups. The statistical optimization scheme allows a prescribed covertness outage probability, whereas the worst-case optimization scheme requires the covertness constraints to hold for all channel errors within the uncertainty set. 

\section{Conclusion}\label{sec:conclusion}
In this paper, we investigated a robust MA-assisted covert communication system with multiple wardens under imperfect CSI.
A weighted max-min SINR maximization problem was formulated by jointly optimizing the transmit beamforming vectors, AN covariance matrix, and MA positions.
Both norm-bounded and statistical warden CSI errors were considered.
Specifically, for norm-bounded CSI errors, we developed a BCD-based algorithm incorporating the S-procedure, SDR, and SCA methods.
For statistical CSI errors, we first derived a tractable deterministic approximation of the covertness outage constraints using the Bernstein-type inequality.
Building upon the proposed BCD optimization framework, we developed an outage-constrained optimization scheme.
Simulation results showed that our proposed method can outperform the FPA schemes in terms of weighted max-min SINR.

\begin{appendices}
\section{\texorpdfstring{Proof of Feasibility of \eqref{SDR_optimal} in Problem \eqref{p2}}{Proof of Feasibility of the Rank-One Recovery}}\label{AppSDR}
Let $\{\overline{\boldsymbol W}_k\}$ and $\overline{\boldsymbol{R}}$ denote a feasible solution of the relaxed SDP for a given $\tau>0$, and define $\overline{\boldsymbol{R}}_{\rm A}\triangleq \overline{\boldsymbol{R}}-\sum_{k\in\mathcal K}\overline{\boldsymbol W}_k$. The constructed rank-one solution is given by $\boldsymbol w_k^\star = (\boldsymbol h_k^H\overline{\boldsymbol W}_k\boldsymbol h_k)^{-1/2}\overline{\boldsymbol W}_k\boldsymbol h_k$ and $\boldsymbol{R}^\star=\overline{\boldsymbol{R}}$.
First, the construction in \eqref{SDR_optimal} yields a rank-one matrix $\boldsymbol W_k^\star=\boldsymbol w_k^\star\boldsymbol w_k^{\star H}$ satisfying
\begin{equation}
\boldsymbol h_k^H\boldsymbol W_k^\star\boldsymbol h_k = \boldsymbol h_k^H\boldsymbol w_k^\star\boldsymbol w_k^{\star H}\boldsymbol h_k = \boldsymbol h_k^H\overline{\boldsymbol W}_k\boldsymbol h_k.
\end{equation}
Thus, the desired signal power in the SINR remains unchanged. Next, we show that $\overline{\boldsymbol W}_k - \boldsymbol W_k^\star \succeq \boldsymbol 0$. For any $\boldsymbol v \in \mathbb C^{M\times 1}$, we have
\begin{equation}
\boldsymbol v^H\!(\overline{\boldsymbol W}\!_{k} \!-\! \boldsymbol W_k^\star)\boldsymbol v \!=\! \boldsymbol v^H\overline{\boldsymbol W}_k\boldsymbol v \!-\! (\boldsymbol h_k^H\overline{\boldsymbol W}_k\boldsymbol h_k)^{-1} |\boldsymbol v^H\overline{\boldsymbol W}_k\boldsymbol h_k|^2.
\end{equation} By the Cauchy--Schwarz inequality, we have
\begin{equation}
(\boldsymbol h_k^H\overline{\boldsymbol W}_k\boldsymbol h_k)(\boldsymbol v^H\overline{\boldsymbol W}_k\boldsymbol v) \ge |\boldsymbol v^H\overline{\boldsymbol W}_k\boldsymbol h_k|^2.
\end{equation}
Therefore, $\boldsymbol v^H(\overline{\boldsymbol W}_k - \boldsymbol W_k^\star)\boldsymbol v \ge 0$ holds for any $\boldsymbol v \in \mathbb C^{M\times 1}$, i.e., $\overline{\boldsymbol W}_k - \boldsymbol W_k^\star \succeq \boldsymbol 0$. We conclude that $\boldsymbol{R}_{\rm A}^\star = \overline{\boldsymbol{R}} - \sum_{k\in\mathcal K}\boldsymbol W_k^\star \succeq \overline{\boldsymbol{R}} - \sum_{k\in\mathcal K}\overline{\boldsymbol W}_k = \overline{\boldsymbol{R}}_{\rm A} \succeq \boldsymbol 0$.
To verify constraint \eqref{Const_p2_sinr}, recall that $\boldsymbol h_k^H\boldsymbol W_k^\star\boldsymbol h_k = \boldsymbol h_k^H\overline{\boldsymbol W}_k\boldsymbol h_k$ and $\boldsymbol{R}^\star = \overline{\boldsymbol{R}}$, which imply $\boldsymbol h_k^H(\boldsymbol{R}^\star - \boldsymbol W_k^\star)\boldsymbol h_k = \boldsymbol h_k^H(\overline{\boldsymbol{R}} - \overline{\boldsymbol W}_k)\boldsymbol h_k$. Thus, we have
\begin{equation}
\frac{\boldsymbol h_k^H\boldsymbol W_k^\star\boldsymbol h_k}{\boldsymbol h_k^H(\boldsymbol{R}^\star - \boldsymbol W_k^\star)\boldsymbol h_k + \sigma_k^2} = \frac{\boldsymbol h_k^H\overline{\boldsymbol W}_k\boldsymbol h_k}{\boldsymbol h_k^H(\overline{\boldsymbol{R}} - \overline{\boldsymbol W}_k)\boldsymbol h_k + \sigma_k^2},
\end{equation}
which confirms that \eqref{Const_p2_sinr} is satisfied.
To verify constraint (\ref{Const_2}), rewrite it as $(1+\eta)\boldsymbol h_q^H\sum_{k\in\mathcal K_c}\boldsymbol W_k\boldsymbol h_q \le \eta(\boldsymbol h_q^H\boldsymbol{R}\boldsymbol h_q + \sigma_q^2)$. Recall that $\overline{\boldsymbol W}_k - \boldsymbol W_k^\star \succeq \boldsymbol 0$ implies $\boldsymbol h_q^H\boldsymbol W_k^\star\boldsymbol h_q \le \boldsymbol h_q^H\overline{\boldsymbol W}_k\boldsymbol h_q$, and $\boldsymbol{R}^\star = \overline{\boldsymbol{R}}$.
Thus, we have
\begin{equation}
\begin{aligned}
  (1+\eta)\boldsymbol h_q^H\sum_{k\in\mathcal K_c}\boldsymbol W_k^\star\boldsymbol h_q &\le (1+\eta)\boldsymbol h_q^H\sum_{k\in\mathcal K_c}\overline{\boldsymbol W}_k\boldsymbol h_q \\
  &\le \eta(\boldsymbol h_q^H\overline{\boldsymbol{R}}\boldsymbol h_q + \sigma_q^2) \\ & = \eta(\boldsymbol h_q^H\boldsymbol{R}^\star\boldsymbol h_q + \sigma_q^2),
\end{aligned}
\end{equation}
which confirms that (\ref{Const_2}) is satisfied. Consequently, all the constraints in the original problem are met.
  
\section{\texorpdfstring{Proof of Lemma~\ref{lem:global_lower_bound}}{Proof of the Global Lower Bound Approximation}}\label{AppGlobalLowerBound}
First, let $\mu\boldsymbol I_N\succeq\boldsymbol M\succeq\nabla^2f(\boldsymbol x)$ for all $\boldsymbol x$.
By applying the generalized form of Lemma~12 in \cite{Sun2017MM}, we have
\begin{equation}\label{eq:lemma_quadratic_upper}
f(\boldsymbol x)
\le{} f(\boldsymbol y)+\nabla f(\boldsymbol y)^T(\boldsymbol x-\boldsymbol y)+\frac{1}{2}(\boldsymbol x-\boldsymbol y)^T\boldsymbol M(\boldsymbol x-\boldsymbol y).
\end{equation}
Then, the above inequality can be rewritten as
\begin{equation}\label{eq:lemma_matrix_lower}
f_2(\boldsymbol x)
\ge{} f_2(\boldsymbol y)+\nabla f_2(\boldsymbol y)^T(\boldsymbol x-\boldsymbol y)-\frac{1}{2}(\boldsymbol x-\boldsymbol y)^T\boldsymbol M(\boldsymbol x-\boldsymbol y),
\end{equation}
where $\boldsymbol M\succeq-\nabla^2f_2(\boldsymbol x)$.
As $\mu\boldsymbol I_N\succeq\boldsymbol M$, we have
\begin{equation}\label{eq:lemma_scalar_lower}
f_2(\boldsymbol x)
\ge{} f_2(\boldsymbol y)+\nabla f_2(\boldsymbol y)^T(\boldsymbol x-\boldsymbol y)-\frac{1}{2}\mu(\boldsymbol x-\boldsymbol y)^T(\boldsymbol x-\boldsymbol y).
\end{equation}
This completes the proof.

\section{Derivation of the Hessian Bounds}\label{AppSINR_a}
The function $a_k(\boldsymbol{x})$ defined in \eqref{eq:ak_position_def} can be rewritten as
\begin{equation}
a_k(\boldsymbol{x}) = 2 \sum_{m\in\mathcal{M}} \sum_{\ell=1}^{L_k} \zeta_{a,k}^{m,\ell} \cos(\kappa_{a,k}^{m,\ell}),
\end{equation}
where $\rho_k^\ell \triangleq \cos(\theta_k^\ell)$, $\kappa_{a,k}^{m,\ell} \triangleq \frac{2\pi}{\lambda}\chi_{k}^{\ell}(x_m)+\angle\alpha_k^*+\angle\beta_k^\ell+\angle w_k^m$, and $\zeta_{a,k}^{m,\ell} \triangleq |\alpha_k^*||\beta_k^\ell||w_k^m|$. Then, we have
\begin{equation}
\begin{aligned}
  &\frac{\partial a_k(\boldsymbol{x})}{\partial x_m} = -\frac{4\pi}{\lambda} \sum_{\ell=1}^{L_k} \zeta_{a,k}^{m,\ell} \rho_k^\ell \sin(\kappa_{a,k}^{m,\ell}),\\
  &\frac{\partial^2 a_k(\boldsymbol{x})}{\partial x_m\partial x_n} = 0,\quad m\ne n,\\
  &\frac{\partial^2 a_k(\boldsymbol{x})}{\partial x_m^2} = -\frac{8\pi^2}{\lambda^2} \sum_{\ell=1}^{L_k} \zeta_{a,k}^{m,\ell} (\rho_k^\ell)^2 \cos(\kappa_{a,k}^{m,\ell}).
\end{aligned}
\end{equation}
Using
$\left|\cos(\kappa_{a,k}^{m,\ell})\right|\!\leq\! 1$ and
$\left|w_k^m\right| \!\leq\! \max_{n\in\mathcal{M}}\left|w_k^n\right|$
, we have
\begin{equation}
\begin{aligned}
\left|\!
\frac{\partial^2 a_k(\boldsymbol{x})}{\partial x_m^2}
\!\right|\!
\!\leq\!
\frac{8\pi^2\!\left|\alpha_k^*\right|\!
}{\lambda^2}
\!\sum_{\ell=1}^{L_k}
\!\left|\beta_k^\ell\right|\!(\rho_k^\ell)^2
\!\max_{n\in\mathcal{M}}\!\left|w_k^n\right|\!
\triangleq\!
\varpi_{a,k},
\end{aligned}
\end{equation}
which implies $\|\nabla^2 a_k(\boldsymbol x)\|_2\le\varpi_{a,k}$.

Following the same approach as for $a_k(\boldsymbol x)$ and using
Theorem~8.1.18 in \cite{HornJohnson}, we further obtain the Hessian bounds
for $b_k(\boldsymbol x)$ and $\varkappa_q(\boldsymbol x)$ as
\begin{equation}
\varpi_{b,k}
=
\frac{16\pi^2 M|\alpha_k|^2}{\lambda^2}
\sum_{\ell=1}^{L_k}\sum_{p=1}^{L_k}|\beta_k^\ell||\beta_k^p|\max_{m,n\in\mathcal{M}}|[\boldsymbol\Xi_k]_{m,n}|,
\end{equation}
where $\|\nabla^2b_k(\boldsymbol x)\|_2\le\max_{m\in\mathcal M}\sum_{n\in\mathcal M}|[\nabla^2b_k(\boldsymbol x)]_{m,n}|\le\varpi_{b,k}$, which also implies $\varpi_{b,k}\boldsymbol{I}_M \succeq \nabla_{\boldsymbol{x}}^2 b_k(\boldsymbol{x})$.
\begin{equation}
\varpi_{\varkappa,q}
=
\frac{16\pi^2 (M-1)}{\lambda^2}\max_{m,n\in\mathcal{M}}|[\bm\Xi]_{m,n}||\beta_q|^2,
\end{equation}
where 
$\varpi_{\varkappa,q}\boldsymbol{I}_M
\succeq
\nabla_{\boldsymbol{x}}^2 \varkappa_q(\boldsymbol{x})$.
\section{\texorpdfstring{Derivation of the Upper Bound in \eqref{eq:mu_upper_two_part_app}}{Derivation of the Covertness Cross Term Upper Bound}}\label{AppCovertSCA}
The cross term $\mu_q(\boldsymbol x,\bm\delta_q)$ in
\eqref{eq:covert_constraint_expanded_position} can be rewritten as
\begin{equation}\label{eq:mu_row_expansion_app}
\mu_q(\boldsymbol{x},\bm\delta_q)
=\sum_{m\in\mathcal{M}}
\left(
\boldsymbol p_{q,m}(x_m)\bm\delta_q
+\bm\delta_q^H\boldsymbol p_{q,m}^H(x_m)
\right),
\end{equation}
where $\boldsymbol p_{q,m}(x_m)\triangleq
\beta_q e^{\jmath\frac{2\pi}{\lambda}\chi_q(x_m)}[\bm\Xi]_{m,:}$
and $[\bm\Xi]_{m,:}$ denotes the $m$-th row of $\bm\Xi$.
The $m$-th element of $\nabla_{\boldsymbol{x}}\mu_q(\boldsymbol{x},\bm\delta_q)$ can be expressed as
\begin{equation}\label{eq:mu_gradient_entry_app}
\frac{\partial \mu_q(\boldsymbol{x},\bm\delta_q)}{\partial x_m}
=
\boldsymbol r_{q,m}(x_m)\bm\delta_q
+
\bm\delta_q^H\boldsymbol r_{q,m}^H(x_m),
\end{equation}
where $\boldsymbol r_{q,m}(x_m)
\triangleq
\jmath \frac{2\pi}{\lambda} \rho_q \boldsymbol p_{q,m}(x_m)$
and $\rho_q = \cos(\theta_q)$.
Then, we have
\begin{equation}
\begin{aligned}
\frac{\partial^2 \mu_q(\boldsymbol{x},\bm\delta_q)}{\partial x_m^2}
&\!=\!-\frac{4\pi^2\rho_q^2}{\lambda^2}
\bigl(
\boldsymbol p_{q,m}(x_m)\bm\delta_q \!+\!\bm\delta_q^H\boldsymbol p_{q,m}^H(x_m)
\bigr),\\
\frac{\partial^2 \mu_q(\boldsymbol{x},\bm\delta_q)}
{\partial x_m\partial x_n}
&\!=\!0,\quad m\ne n.
\end{aligned}
\end{equation}
By applying the triangle and Cauchy--Schwarz inequalities and using
$\|\bm\delta_q\|_2 \!\leq\! e_q$ and $\left\|[\bm\Xi]_{m,:}\right\|_2 \!\leq\!\max_{n\in\mathcal{M}} \left\|[\bm\Xi]_{n,:}\right\|_2 $, we have
\begin{equation}
\left|
\frac{\partial^2 \mu_q(\boldsymbol{x},\bm\delta_q)}{\partial x_m^2}
\right|
\le
\frac{8\pi^2}{\lambda^2}\rho_q^2|\beta_q|e_q
\max_{m\in\mathcal{M}}
\left\|
[\bm\Xi]_{m,:}
\right\|_2 \triangleq \varpi_{\mu,q},
\end{equation}
where
$ \varpi_{\mu,q}\mathbf{I}_M
\succeq
\nabla_{\boldsymbol{x}}^2\mu_q(\boldsymbol{x},\bm\delta_q) $.

\section{\texorpdfstring{Proof of Proposition~\ref{prop:bcd_wc_monotonicity}}{Proof of Proposition 1}}\label{AppBCDWC}
Let $F$ and $\overline F$ denote the weighted max-min SINRs of the rank-one and relaxed solutions, respectively. First, in steps~2--4 of Algorithm~\ref{alg:robust_bcd}, for given $\boldsymbol x^{(r)}$, the covariances are optimized and rank-one solutions are recovered.
Secondly, in steps~5--6, for given $\{\boldsymbol w_k^{(r+1)}\}$ and $\boldsymbol R_{\rm A}^{(r+1)}$, the antenna positions are updated by SCA. Combining these two steps, we have~\cite{HaoFluidISAC2025}
\begin{equation}\label{eq:bcd_wc_outer_monotonicity}
\begin{aligned}
F\bigl(\boldsymbol w_k^{(r)},\boldsymbol R_{\rm A}^{(r)},\boldsymbol x^{(r)}\bigr)
&\!\overset{(a)}{\le}\!
\overline F\bigl(\overline{\boldsymbol W}_k^{(r+1)},\overline{\boldsymbol R}^{(r+1)},\boldsymbol x^{(r)}\bigr)\\
&\!\overset{(b)}{=}\!
F\bigl(\boldsymbol w_k^{(r+1)},\boldsymbol R_{\rm A}^{(r+1)},\boldsymbol x^{(r)}\bigr)\\
&\!\overset{(c)}{\le}\!
F\bigl(\boldsymbol w_k^{(r+1)},\boldsymbol R_{\rm A}^{(r+1)},\boldsymbol x^{(r+1)}\bigr),
\end{aligned}
\end{equation}
where (a) follows from initializing bisection for \eqref{p2} with the current feasible lower bound, and (b) holds because the rank-one recovery in \eqref{SDR_optimal} preserves each user's SINR at $\boldsymbol x^{(r)}$, as proved in Appendix~\ref{AppSDR}. For (c), \eqref{eq:alpha_update_position} ensures tight SINR lower bounds at each SCA iteration. The optimal value of \eqref{p3_sca} lies between the current and updated $F$. Both sequences are non-decreasing and bounded above by \eqref{Const_3}--\eqref{Const_5}. Thus, the convergence of Algorithm~\ref{alg:robust_bcd} can be guaranteed.

\end{appendices}

\bibliographystyle{IEEEtran}
\bibliography{reference.bib}


\end{document}